\documentclass[english]{article}
\usepackage[T1]{fontenc}
\usepackage[latin9]{inputenc}
\usepackage{color}
\usepackage{babel}
\usepackage{mathrsfs}
\usepackage{amsmath}
\usepackage{amssymb}
\usepackage{graphicx}
\usepackage{setspace}
\usepackage[unicode=true,pdfusetitle,
 bookmarks=true,bookmarksnumbered=false,bookmarksopen=false,
 breaklinks=false,pdfborder={0 0 1},backref=false,colorlinks=false]
 {hyperref}
\usepackage{breakurl}

\makeatletter

\providecommand{\tabularnewline}{\\}

\usepackage{cite}

\usepackage{graphicx}
\DeclareMathOperator*{\name}{\raisebox{-0.8ex}{\scalebox{2.5}{$\ast$}}}

\makeatother

\begin{document}
\noindent \begin{flushleft}
\textbf{\Large{}Mathematical studies of the dynamics of finite-size
binary neural networks: A review of recent progress}\\
\par\end{flushleft}{\Large \par}

\noindent \begin{flushleft}
Diego Fasoli$^{1,\ast}$, Stefano Panzeri$^{1}$ 
\par\end{flushleft}

\medskip{}
\noindent \begin{flushleft}
\textbf{{1} Laboratory of Neural Computation, Center for Neuroscience
and Cognitive Systems @UniTn, Istituto Italiano di Tecnologia, 38068
Rovereto, Italy}
\par\end{flushleft}

\noindent \begin{flushleft}
\textbf{$\ast$ Corresponding Author. E-mail: diego.fasoli@iit.it}
\par\end{flushleft}

\section*{Abstract}

\noindent Traditional mathematical approaches to studying analytically
the dynamics of neural networks rely on the mean-field approximation,
which is rigorously applicable only to networks of infinite size.
However, all existing real biological networks have finite size, and
many of them, such as microscopic circuits in invertebrates, are composed
only of a few tens of neurons. Thus, it is important to be able to
extend to small-size networks our ability to study analytically neural
dynamics. Analytical solutions of the dynamics of finite-size neural
networks have remained elusive for many decades, because the powerful
methods of statistical analysis, such as the central limit theorem
and the law of large numbers, do not apply to small networks. In this
article, we critically review recent progress on the study of the
dynamics of small networks composed of binary neurons. In particular,
we review the mathematical techniques we developed for studying the
bifurcations of the network dynamics, the dualism between neural activity
and membrane potentials, cross-neuron correlations, and pattern storage
in stochastic networks. Finally, we highlight key challenges that
remain open, future directions for further progress, and possible
implications of our results for neuroscience.

\section{Introduction \label{sec:Introduction}}

Understanding the dynamics of networks of neurons, and of how such
networks represent, process and exchange information by means of the
temporal evolution of their activity, is one of the central problems
in neuroscience. Real networks of neurons are highly complex both
in terms of structure and physiology. Introducing details of this
complexity greatly complicates the tractability of the models. Thus,
a mathematical model of neural networks needs to be carefully designed,
by finding a compromise between the elements of biological complexity
and plausibility that are introduced, and the analytical tractability
of the resulting model \cite{Lytton2002}.

A wide set of mathematical models have been proposed to investigate
the behavior of biological neural networks \cite{Ermentrout1998,Izhikevich2007,Ashwin2016}.
Typically, these models attempt to simplify as much as possible the
original system they describe, without losing the properties that
give rise to the most interesting emergent phenomena observed in biological
systems. Binary neural network models \cite{Amari1972,Little1974,Hopfield1982,Coolen2001a,Coolen2001b}
represent one of the most successful examples in finding a good compromise
between keeping simplicity to enhance tractability, while yet achieving
a rich dynamics with a set of complex emergent network properties.

Binary models describe the dynamical properties of networks composed
of threshold units, which integrate their inputs to produce a binary
output, namely a high (respectively low) output firing rate when their
membrane potential does (respectively does not) exceed a given threshold
(see Sec.~(\ref{sec:The-binary-network-model}) for more details).
Among the wide set of neural network models proposed by computational
neuroscientists, threshold units represent one of the most convenient
tools for studying the dynamical and statistical properties of neural
circuits. The relative ease with which these models can be investigated
analytically, is a consequence of their thresholding activation function,
which can be considered as the simplest, piecewise-constant, approximation
of the non-linear (and typically sigmoidal shaped) graded input-output
relationship of biological neurons. Despite their simplicity, as shown
both by classic work \cite{Amari1972,Little1974,Hopfield1982,Peretto1984,Ginzburg1994},
as well as by our work reviewed here \cite{Fasoli2018a,Fasoli2018b,Fasoli2019},
the jump discontinuity of their activation function at the threshold
is sufficient to endow binary networks with a complex set of useful
emergent dynamical properties and non-linear phenomena, such as attractor
dynamics \cite{Amit1989}, formation of patterns and oscillatory waves
\cite{Amari1977}, chaos \cite{vanVreeswijk1996}, and information
processing capabilities \cite{Amari1975}, which are reminiscent of
neuronal activity in biological networks.

The importance of binary network models is further strengthened by
their close relationship with spin networks studied in physics \cite{Sherrington1976,Kirkpatrick1978,Mezard1984,Mezard1986}.
The temporal evolution of a binary network in the zero-noise limit
is isomorphic to the dynamics of kinetic models of spin networks at
absolute temperature \cite{Glauber1963}. This allowed computational
neuroscientists to study the behavior of large-size binary networks,
by applying the powerful techniques of statistical mechanics already
developed for spin models (see e.g. \cite{Coolen1993}).

Sizes of brains and of specialized neural networks within brains change
considerably across animal species, ranging from few tens of neurons
in invertebrates such as rotifers and nematodes, to billions of neurons
in cetaceans and primates \cite{Williams1988}. Network size changes
also across levels of spatial organizations, ranging from microscopic
and mesoscopic levels of organization in cortical micro-columns and
columns (including from few tens \cite{Mountcastle1997} to few tens
of thousands of neurons \cite{Helmstaedter2007,Meyer2010}), to several
orders of magnitude more in large macroscopic networks, such as the
resting state networks in the human brain, that involve many brain
areas \cite{DeLuca2006,Mantini2007}. For this reason, it is important
to be able to study mathematically the dynamics of binary neural network
models (or of any network model, see e.g. \cite{Beer1995,Pasemann2002,Fasoli2016}),
for a wide range of different network sizes.

Large-scale networks composed of several thousands of neurons or more,
are typically studied by taking advantage of the powerful techniques
of statistical mechanics, such as the law of large numbers and the
central limit theorem, see e.g. \cite{Coolen1993,Ginzburg1994,vanVreeswijk1998,Fasoli2018a}.
These theories, such as those developed in physics for spin models,
typically approximate the interaction of all the other neurons to
a given neuron with a mean field, namely an effective interaction
which is obtained by averaging the interactions of the other neurons.
This allows one to dimensionally reduce the model, by transforming
the set of equations of a large network into a single-neuron equation.
Therefore the mean-field theories represent a powerful tool for gaining
insight into the behavior of large networks, at a relatively low cost.

However, statistical mechanics does not apply to small-scale networks
containing only few tens of neurons, which therefore prove much more
difficult to study mathematically. Importantly, several studies, see
e.g. \cite{Cessac1995,vanVreeswijk1998,Renart2010,Fasoli2016,Fasoli2018a,Fasoli2018b},
have shown that the dynamics of both binary and graded neural networks
in the large-size limit can be qualitatively, not only quantitatively,
very different from that of the same model with small or finite size.
For this reason, the computational investigation of neural networks
composed of a small number of threshold units requires the development
of new specialized analytical and numerical techniques. Recently,
we proposed such techniques \cite{Fasoli2018a,Fasoli2018b,Fasoli2019},
which we critically review in this paper.

In Sec.\textcolor{blue}{~}(\ref{sec:The-binary-network-model}) we
introduce the binary network model that we analyze in this review,
while in Sec.\textcolor{blue}{~}(\ref{sec:Analysis-of-bifurcations-in-deterministic-networks})
we focus specifically on the zero-noise limit of small networks, and
we characterize mathematically the bifurcation points of the network
dynamics in single network realizations. In Sec.\textcolor{blue}{~}(\ref{sec:Stochastic-networks})
we describe the techniques that we developed for studying small networks
when external sources of noise are added to the neural equations.
In particular, in SubSec.\textcolor{blue}{~}(\ref{subsec:Dualism-between-neural-activity-and-membrane-potentials})
we describe the dualism between neural activity and membrane potentials,
and we derive a complete description of the probabilistic behavior
of the network in the long-time limit. In SubSecs.\textcolor{blue}{~}(\ref{subsec:Cross-neuron-correlations})
and (\ref{subsec:Pattern-storage-and-retrieval-in-presence-of-noise}),
we introduce, under some assumptions on the nature of the noise sources,
exact analytical expressions of the cross-neuron correlations, and
a learning rule for storing patterns and sequences of neural activity.
Then, in Sec.\textcolor{blue}{~}(\ref{sec:Networks-with-quenched-disorder}),
we extend the study of bifurcations of Sec.\textcolor{blue}{~}(\ref{sec:Analysis-of-bifurcations-in-deterministic-networks})
to networks with quenched disorder across multiple network realizations.
To conclude, in Sec.\textcolor{blue}{~}(\ref{sec:Discussion}) we
discuss the advantages and weaknesses of our techniques. Moreover,
we highlight key challenges that remain open, as well as future directions
for further progress in the mathematical study of binary networks,
and possible implications of our results for neuroscience.

\section{The binary network model \label{sec:The-binary-network-model}}

In this review we assume that neural activity evolves in discrete
time steps, and that the threshold units are synchronously updated.
These assumptions are often used in studying network dynamics, see
e.g. \cite{Amari1972,Little1974,Peretto1984,Ermentrout1998}. A threshold
unit, or \textit{artificial neuron}, is a logic gate or a mathematical
function that mimics the working mechanisms of a biological neuron.
Typically the unit receives several inputs, which can be loosely interpreted
as postsynaptic potentials at neural dendrites, and sums them to produce
a binary or digital output (also known as \textit{activation} or \textit{neural
activity}). Usually, each input of a threshold unit is multiplied
by a so called \textit{synaptic weight}, which represents the strength
of connections between pairs of neurons. Moreover, the sum of the
weighted inputs is passed through a piecewise-constant (or Heaviside)
thresholding function, also known as \textit{activation function}
or \textit{transfer function}. If the sum of the weighted inputs exceeds
a threshold, the output is set to one, and the artificial neuron is
said to fire at that rate. On the contrary, if the sum is below the
threshold, the output is set to zero, and the neuron is quiescent.
For this reason, the binary output of a threshold unit can be loosely
interpreted as the firing rate of the postsynaptic neuron, namely
as the number of spikes per second of its action potential, which
propagates along its axon toward other neurons in the network.

The basic set of equations defining the dynamics of the discrete-time
binary network is:

\begin{spacing}{0.8}
\begin{center}
{\small{}
\begin{equation}
A_{i}\left(t+1\right)=\mathscr{H}\left(\sum_{j=0}^{N-1}J_{i,j}A_{j}\left(t\right)+\mathfrak{I}_{i}+\sum_{j=0}^{N-1}\sigma_{i,j}^{\mathcal{N}}\mathcal{N}_{j}\left(t\right)-\theta_{i}\right),\quad i=0,...,N-1,\label{eq:activity-based-equations}
\end{equation}
}
\par\end{center}{\small \par}
\end{spacing}

\noindent which describes the temporal evolution of the neural activity
$A_{i}$ of the $i$th neuron, from the time instant $t$ to the time
instant $t+1$. In Eq.\textcolor{blue}{~}(\ref{eq:activity-based-equations}),
$N$ represents the size (namely the number of threshold units) of
the network. The matrix $J=\left[J_{i,j}\right]_{i,j=0,\cdots,N-1}$
is the (generally asymmetric) synaptic connectivity matrix of the
network, whose entries $J_{i,j}$ are time-independent and represent
the strength or weight of the synaptic connection from the $j$th
(presynaptic) neuron to the $i$th (postsynaptic) neuron. Moreover,
$\mathfrak{I}_{i}+\sum_{j=0}^{N-1}\sigma_{i,j}^{\mathcal{N}}\mathcal{N}_{j}\left(t\right)$
represents the total external input current (i.e. the stimulus) to
the $i$th neuron. In more detail, $\mathfrak{I}_{i}$ is the time-independent
deterministic component of the stimulus, while its stochastic component
is the sum of $N$ random variables $\sigma_{i,j}^{\mathcal{N}}\mathcal{N}_{j}\left(t\right)$,
each one having zero mean and standard deviation $\sigma_{i,j}^{\mathcal{N}}$.
The vector $\boldsymbol{\mathcal{N}}\overset{\mathrm{def}}{=}\left[\begin{array}[t]{ccc}
\mathcal{N}_{0}, & \ldots, & \mathcal{N}_{N-1}\end{array}\right]^{T}$ represents a collection of stochastic variables with unit standard
deviation, whose joint probability distribution $p_{\boldsymbol{\mathcal{N}}}$
is arbitrary. Then, in Eq.\textcolor{blue}{~}(\ref{eq:activity-based-equations}),
$\mathscr{H}\left(\cdot\right)$ is the Heaviside activation function
with threshold $\theta$, which is defined as follows:
\begin{spacing}{0.8}
\begin{center}
{\small{}
\[
\mathscr{H}\left(x-\theta\right)=\begin{cases}
0, & \mathrm{if}\;\;x<\theta\\
\\
1, & \mathrm{if}\;\;x\geq\theta.
\end{cases}
\]
}
\par\end{center}{\small \par}
\end{spacing}

\noindent It is important to observe that, unlike the classic Hopfield
network \cite{Hopfield1982}, which is symmetric and asynchronously
updated, a Lyapunov function for synchronous networks with asymmetric
synaptic weights, like ours, is generally not known. For this reason,
the analytical investigation of the network dynamics determined by
Eq.\textcolor{blue}{~}(\ref{eq:activity-based-equations}) proves
much more challenging. In Secs.\textcolor{blue}{~}(\ref{sec:Analysis-of-bifurcations-in-deterministic-networks})-(\ref{sec:Networks-with-quenched-disorder}),
we will review the techniques, that we developed in \cite{Fasoli2018a,Fasoli2018b,Fasoli2019},
for investigating the dynamical and probabilistic properties of Eq.\textcolor{blue}{~}(\ref{eq:activity-based-equations})
in small networks.

\section{Analysis of bifurcations in deterministic networks \label{sec:Analysis-of-bifurcations-in-deterministic-networks}}

An important problem in the theory of binary networks is represented
by the study of the qualitative changes in the dynamics of their neuronal
activity, which typically are elicited by variations in the external
stimuli. These changes of dynamics are named in mathematical terms
as \textit{bifurcations} \cite{Kuznetsov1998}. Seminal work in physics
focused on the study of bifurcations in infinite-size spin networks,
see e.g. \cite{Sherrington1976,DeAlmeida1978,Mezard1984}. On the
other hand, the theory of bifurcations of non-smooth dynamical systems
composed of a finite number of units, including those with discontinuous
functions such as binary networks, has been developed mostly for continuous-time
models, see e.g. \cite{Leine2000,Awrejcewicz2003,Leine2006,Makarenkov2012,Harris2015},
and for piecewise-smooth continuous maps \cite{Parui2002}. Despite
the importance of discontinuous maps in computational neuroscience,
the development of new techniques for studying their bifurcation structure
has received much less attention \cite{Avrutin2006}.

In \cite{Fasoli2018a,Fasoli2019} we tackled the problem of deriving
the bifurcations in the dynamics of the neural activity, for finite-size
binary networks with arbitrary connectivity matrix. As is common practice,
we performed the bifurcation analysis in the zero-noise limit, i.e.
for $\sigma_{i,j}^{\mathcal{N}}\rightarrow0\;\forall i,j$ (see Eq.~(\ref{eq:activity-based-equations})).
In particular, we studied how the dynamics of neural activity switches
between stationary states and neural oscillations, when varying the
external stimulus to the network. Because of the discrete nature of
the neural activity, there exists only a finite number of stationary
and oscillatory solutions to Eq.~(\ref{eq:activity-based-equations}).
This allowed us to introduced a combinatorial brute-force approach
for studying the bifurcation structure of binary networks, which we
describe briefly below.

We introduce the vector $\boldsymbol{A}\overset{\mathrm{def}}{=}\left[\begin{array}[t]{ccc}
A_{0}, & \ldots, & A_{N-1}\end{array}\right]^{T}$ containing the activities of all $N$ neuron, and the sequence $\mathcal{S}\left(0,\mathcal{T}\right)$
of activity vectors $\boldsymbol{A}^{\left(0\right)}\rightarrow\boldsymbol{A}^{\left(1\right)}\rightarrow\cdots\rightarrow\boldsymbol{A}^{\left(\mathcal{T}\right)}$,
for some $1\leq\mathcal{T}\leq2^{N}$. Given a network with $\mathfrak{P}$
distinct input currents $I_{0},\cdots,I_{\mathfrak{P}-1}$, we also
define $\Gamma_{I_{\alpha}}$ to be the set of neurons that share
the same external current $I_{\alpha}$ (namely $\Gamma_{I_{\alpha}}\overset{\mathrm{def}}{=}\left\{ i\in\left\{ 0,\cdots,N-1\right\} :\;\mathfrak{I}_{i}=I_{\alpha}\right\} $),
and $\Gamma_{I_{\alpha},x}^{\left(j\right)}\overset{\mathrm{def}}{=}\left\{ i\in\Gamma_{I_{\alpha}}:\;A_{i}^{\left(j\right)}=x\right\} $
for $x\in\left\{ 0,1\right\} $. Then, in \cite{Fasoli2019} we proved
that the sequence $\mathcal{S}\left(0,\mathcal{T}\right)$ is a solution
of Eq.~(\ref{eq:activity-based-equations}) in the time range $\left[0,\mathcal{T}\right]$
(i.e. $\boldsymbol{A}\left(t=j\right)=\boldsymbol{A}^{\left(j\right)}$
for $j=0,\cdots,\mathcal{T}$), for every combination of stimuli $\left(I_{0},\cdots,I_{\mathfrak{P}-1}\right)\in\mathfrak{V}=\mathcal{V}_{0}\times\cdots\times\mathcal{V}_{\mathfrak{P}-1}$,
where:

{\small{}
\begin{align}
 & \mathcal{V}_{\alpha}\overset{\mathrm{def}}{=}\begin{cases}
\left(-\infty,\Xi_{\alpha}\right) & \mathrm{if}\;\;\Gamma_{I_{\alpha},1}^{\left(j+1\right)}=\emptyset\;\;\forall j\in\mathscr{T}\\
\\
\left[\Lambda_{\alpha},+\infty\right) & \mathrm{if}\;\;\Gamma_{I_{\alpha},0}^{\left(j+1\right)}=\emptyset\;\;\forall j\in\mathscr{T}\\
\\
\left[\Lambda_{\alpha},\Xi_{\alpha}\right) & \mathrm{otherwise}
\end{cases}\nonumber \\
\nonumber \\
 & \mathscr{T}\overset{\mathrm{def}}{=}\left\{ 0,\cdots,\mathcal{T}-1\right\} ,\quad\mathscr{T}_{\alpha,x}\overset{\mathrm{def}}{=}\left\{ j\in\mathscr{T}:\;\Gamma_{I_{\alpha},x}^{\left(j+1\right)}\neq\emptyset\right\} \label{eq:neural-sequence-stimulus-range}\\
\nonumber \\
 & \Lambda_{\alpha}\overset{\mathrm{def}}{=}\underset{j\in\mathscr{T}_{\alpha,1}}{\max}\left(\underset{i\in\Gamma_{I_{\alpha},1}^{\left(j+1\right)}}{\max}\mathcal{I}_{i}^{\left(j\right)}\right),\quad\Xi_{\alpha}\overset{\mathrm{def}}{=}\underset{j\in\mathscr{T}_{\alpha,0}}{\min}\left(\underset{i\in\Gamma_{I_{\alpha},0}^{\left(j+1\right)}}{\min}\mathcal{I}_{i}^{\left(j\right)}\right)\nonumber \\
\nonumber \\
 & \mathcal{I}_{i}^{\left(j\right)}\overset{\mathrm{def}}{=}\theta_{i}-\sum_{k=0}^{N-1}J_{i,k}A_{k}^{\left(j\right)}.\nonumber 
\end{align}
}{\small \par}

\noindent A neural sequence loses its stability, turning into another
sequence, at the boundaries $\Lambda_{\alpha}$ and $\Xi_{\alpha}$,
which therefore represent the coordinates of the bifurcation points
of the neural activity. In this review, we focus specifically on the
subset of sequences that satisfy the additional constraint $\boldsymbol{A}^{\left(0\right)}=\boldsymbol{A}^{\left(\mathcal{T}\right)}$:
these sequences represent the candidate oscillatory solutions with
period $\mathcal{T}$ of Eq.~(\ref{eq:activity-based-equations}).
We also observe that, in the special case $\mathcal{T}=1$, we obtain
the set of candidate stationary solutions of the network equations.
For this reason, the bifurcation diagram of a binary network can be
decomposed into two panels, the \textit{oscillation} and the \textit{multistability
diagrams}. These diagrams describe the relationship between the oscillatory/stationary
solutions of Eq.~(\ref{eq:activity-based-equations}), and the set
of stimuli. In other words, these diagrams display the fragmentation
of the stimulus space into areas where several oscillatory solutions
occur, and/or where the network is (multi)stable.

It is important to note that only the sequences whose hyperrectangles
$\mathfrak{V}$ have positive hypervolumes (i.e. the sequences that
satisfy the condition $\Lambda_{\alpha}<\Xi_{\alpha}$, for every
$\alpha$ and $j$ such that $\Gamma_{I_{\alpha},0}^{\left(j+1\right)},\;\Gamma_{I_{\alpha},1}^{\left(j+1\right)}\neq\emptyset$)
are solutions of Eq.~(\ref{eq:activity-based-equations}), for some
combinations of stimuli. On the contrary, if the hypervolume of $\mathfrak{V}$
is zero, the corresponding neural sequence is never a solution of
Eq.~(\ref{eq:activity-based-equations}). Unfortunately, the sequences
with positive hypervolumes are not known a priori, therefore they
must be found through a brute-force searching procedure. Because of
the combinatorial explosion of the number of possible sequences for
increasing $N$, typically brute-force algorithms have at least exponential
complexity with respect to the network size. Therefore they can be
applied only to small networks (typically $N<30$), regardless of
the density of their synaptic connections. However, real cortical
circuits are typically very sparse \cite{Kotter2003}, therefore in
\cite{Fasoli2019} we developed an efficient algorithm specifically
designed for networks with a low density of the synaptic connections.
This efficient algorithm takes advantage of the information provided
by the absence of the synaptic connections among the threshold units
to speed up the detection of the oscillatory and stationary solutions
of Eq.~(\ref{eq:activity-based-equations}). In other words, the
sparse-efficient algorithm avoids checking the sequences of neural
activity vectors that are not compatible with the topology of the
synaptic connections, resulting in a much faster calculation of the
bifurcation structure of the network model. The interested reader
is referred to \cite{Fasoli2019} for a detailed discussion of the
algorithm.

\begin{figure}
\begin{centering}
\includegraphics[scale=0.31]{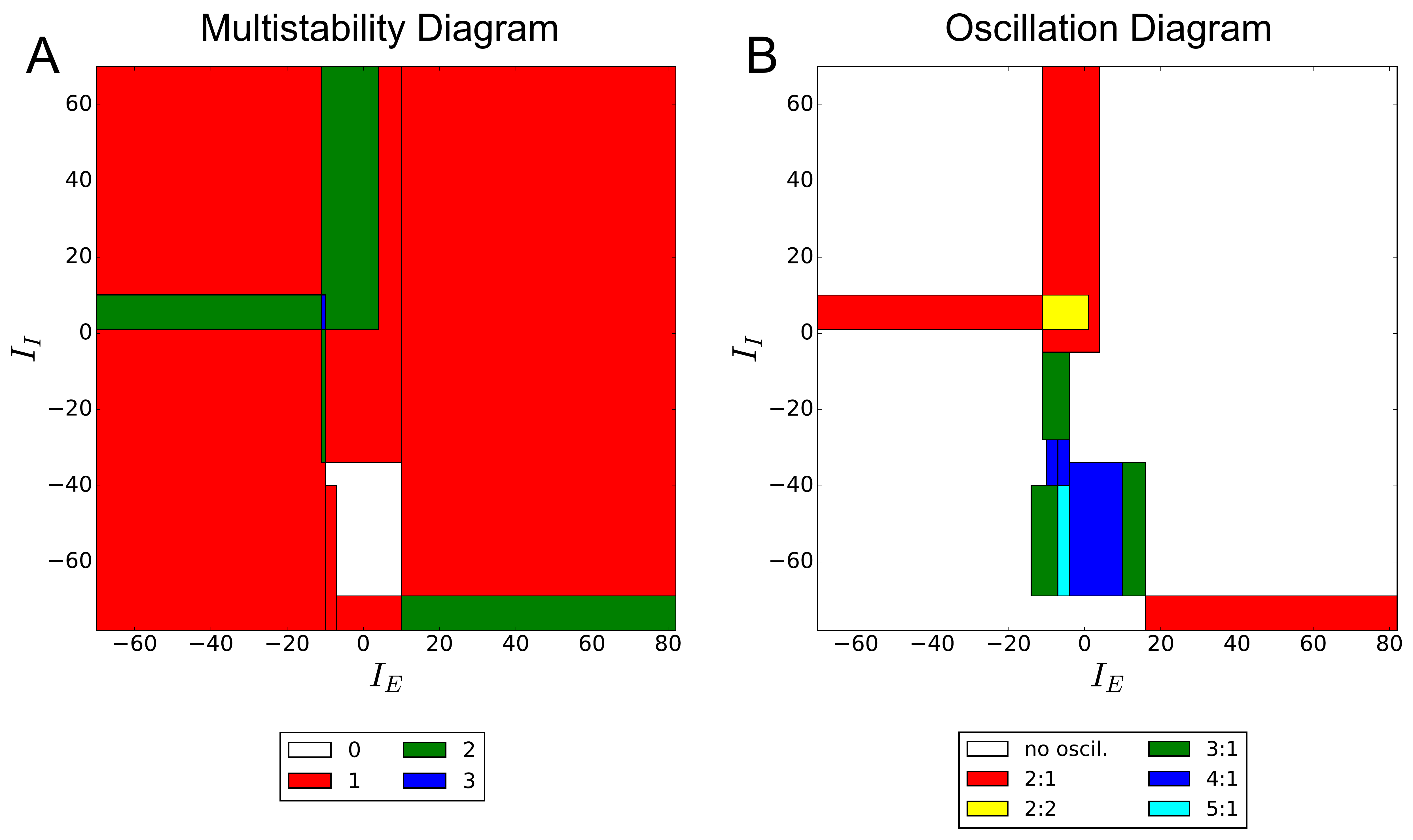}
\par\end{centering}
\caption{\label{fig:bifurcation-diagram} \small\textbf{ An example of bifurcation
diagram.} This figure shows the bifurcation diagram of the binary
network, obtained for the parameters in Tab.~(\ref{tab:Parameters-1}).
We supposed that the excitatory neurons with indexes $i=0$ and $i=1$
receive an arbitrary external stimulus $\mathfrak{I}_{0}=\mathfrak{I}_{1}=I_{E}$,
which represents the first bifurcation parameter, while the excitatory
neuron with index $i=2$ receives a fixed stimulus $\mathfrak{I}_{2}=10$.
Moreover, we assumed that the inhibitory neuron with index $i=3$
receives an arbitrary stimulus $\mathfrak{I}_{3}=I_{I}$, which represents
the second bifurcation parameter, while the inhibitory neuron with
index $i=4$ receives a fixed stimulus $\mathfrak{I}_{4}=5$. Then,
we plotted the multistability and oscillation diagrams in the $I_{E}-I_{I}$
plane, according to Eq.~(\ref{eq:neural-sequence-stimulus-range}).
A) Multistability diagram. Each color represents a different degree
of multistability (white = astable, red = monostable, green = bistable,
blue = tristable). B) Oscillation diagram. Each color represents a
different set of oscillatory solutions of Eq.~(\ref{eq:activity-based-equations})
(the notation $x:y$ reveals the formation of $y$ distinct oscillations
with period $\mathcal{T}=x$). For example, for every combination
of stimuli $\left(I_{E},I_{I}\right)$ that lies in the yellow area,
Eq.~(\ref{eq:activity-based-equations}) has $2$ oscillatory solutions
with period $\mathcal{T}=2$, while for every combination in the green
areas, the equation has an oscillatory solution with period $\mathcal{T}=3$.
Note that, for other values of the network parameters, oscillations
with distinct periods may coexist in the same area.}
\end{figure}

\begin{figure}
\begin{centering}
\includegraphics[scale=0.35]{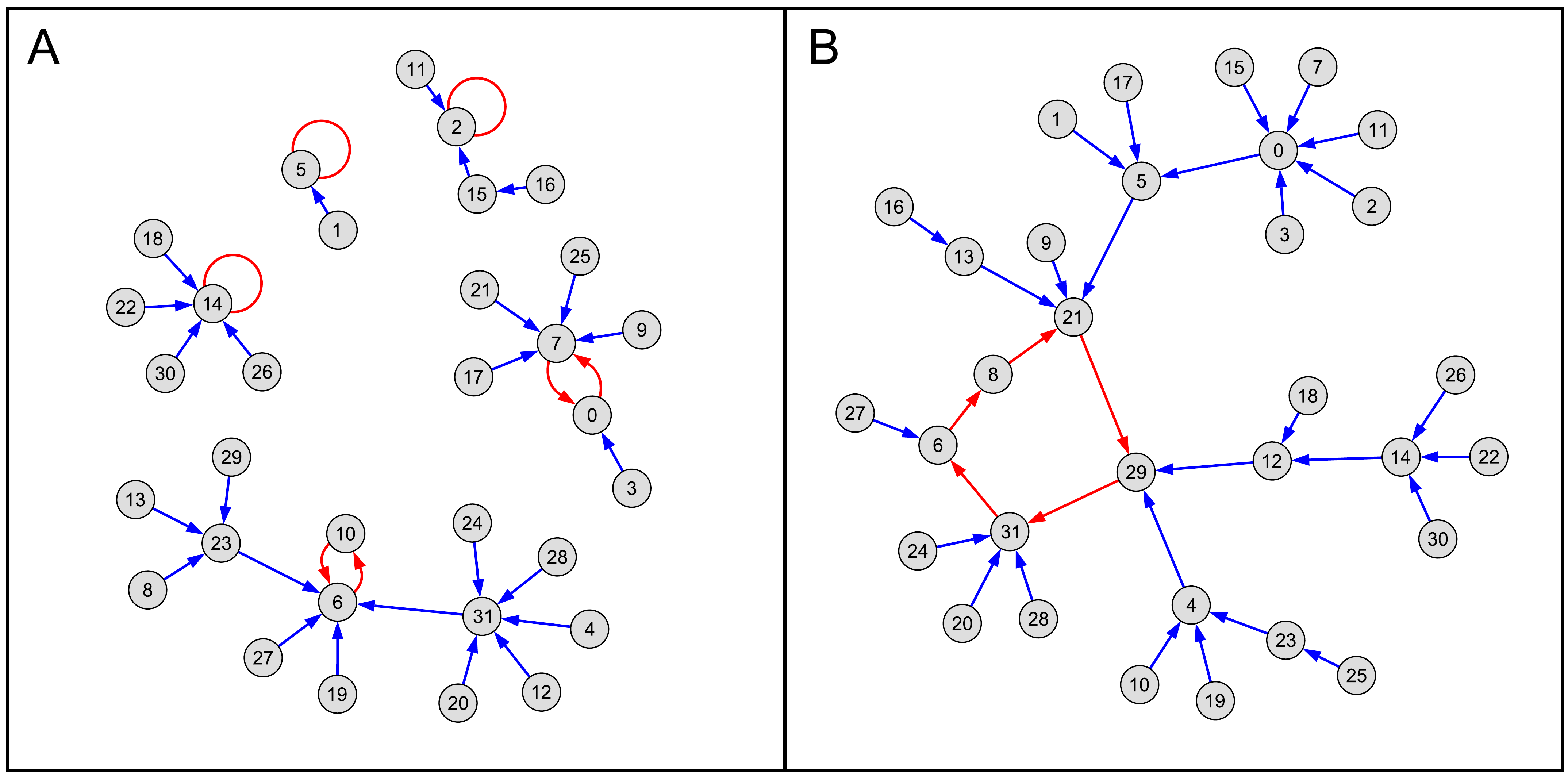}
\par\end{centering}
\caption{\label{fig:graphs} \small\textbf{ Examples of state-to-state transitions
of a binary network.} This figure shows the allowed transitions between
states of neural activity, obtained for the network parameters in
Tab.~(\ref{tab:Parameters-1}) and $\mathfrak{I}_{2}=10$, $\mathfrak{I}_{4}=5$.
The nodes in the graphs represent the $2^{N}$ states of the neural
activity vector $\boldsymbol{A}$ (e.g. the node $26$ corresponds
to the state $\boldsymbol{A}=\left[\protect\begin{array}[t]{ccccc}
1, & 1, & 0, & 1, & 0\protect\end{array}\right]^{T}$), while the arrows represent the allowed transitions between these
states. A) State-to-state transitions, obtained for $I_{E}=-10.5$
and $I_{I}=6$. We highlighted in red $3$ stationary states (i.e.
the nodes $2$, $5$ and $14$) and $2$ oscillations of period $\mathcal{T}=2$
(i.e. $0\rightarrow7\rightarrow0$ and $6\rightarrow10\rightarrow6$).
Note that, as expected, the point in the stimulus plane with coordinates
$\left(I_{E},I_{I}\right)=\left(-10.5,6\right)$ lies in the blue
area of the multistability diagram (see Fig.~(\ref{fig:bifurcation-diagram}),
panel A), which corresponds to tristability, and in the yellow area
of the oscillation diagram (see Fig.~(\ref{fig:bifurcation-diagram}),
panel B), which corresponds to the formation of $2$ oscillations
with period $\mathcal{T}=2$. B) State-to-state transitions, obtained
for $I_{E}=-5$ and $I_{I}=-55$. We highlighted in red an oscillations
of period $\mathcal{T}=5$ (i.e. $6\rightarrow8\rightarrow21\rightarrow29\rightarrow31\rightarrow6$).
Note that the point in the stimulus plane with coordinates $\left(I_{E},I_{I}\right)=\left(-5,-55\right)$
lies in the white area of the multistability diagram, where Eq.~(\ref{eq:activity-based-equations})
has no stationary solutions, and in the cyan area of the oscillation
diagram, which corresponds to the formation of an oscillation with
period $\mathcal{T}=5$.}
\end{figure}

In Fig.~(\ref{fig:bifurcation-diagram}) we show an example of bifurcation
diagram, that we obtained from Eq.~(\ref{eq:neural-sequence-stimulus-range}),
in the specific case of the network parameters reported in Tab.~(\ref{tab:Parameters-1}).
\begin{table}
\begin{centering}
\textbf{\footnotesize{}}%
\begin{tabular}{|c|}
\hline 
\tabularnewline
{\footnotesize{}$J=\left[\begin{array}{ccccc}
0 & 17 & 17 & -43 & -6\\
25 & 0 & 15 & -3 & -32\\
10 & 1 & 0 & -10 & -7\\
50 & 29 & 6 & 0 & -15\\
7 & 28 & 5 & -95 & 0
\end{array}\right],\quad\theta_{0}=\cdots=\theta_{4}=1$}\tabularnewline
\tabularnewline
\hline 
\end{tabular}
\par\end{centering}{\footnotesize \par}
\caption{\label{tab:Parameters-1} \textbf{Network parameters 1}. This table
reports the values of the network parameters that we used for plotting
Figs.~(\ref{fig:bifurcation-diagram}) - (\ref{fig:correlation}).}
\end{table}
 In this example, we consider a network with heterogeneous random
synaptic weights, which is composed of $3$ excitatory neurons and
$2$ inhibitory neurons. Moreover, in Fig.~(\ref{fig:graphs}), we
show two examples of state-to-state transitions, obtained by solving
Eq.~(\ref{eq:activity-based-equations}) in the zero-noise limit,
for all the $2^{N}$ initial conditions of the network dynamics (i.e.
from $\boldsymbol{A}\left(t=0\right)=\left[\begin{array}[t]{ccc}
0, & \ldots, & 0\end{array}\right]^{T}$ to $\boldsymbol{A}\left(t=0\right)=\left[\begin{array}[t]{ccc}
1, & \ldots, & 1\end{array}\right]^{T}$). 

\section{Stochastic networks \label{sec:Stochastic-networks}}

\subsection{Dualism between neural activity and membrane potentials \label{subsec:Dualism-between-neural-activity-and-membrane-potentials}}

\noindent In this section we show the existence of a dualism between
the neural activity states and the membrane potentials in a network
composed of binary neurons. These variables are intrinsically related,
but have distinct dynamical aspects.

As we explained in Sec.\textcolor{blue}{~}(\ref{sec:The-binary-network-model}),
the term $\sum_{j=0}^{N-1}J_{i,j}A_{j}\left(t\right)$ in Eq.\textcolor{blue}{~}(\ref{eq:activity-based-equations})
can be loosely interpreted as the weighted sum of postsynaptic potentials
at neural dendrites. Therefore this term, plus the eventual external
stimulus to the $i$th neuron, can be interpreted as the total membrane
potential $V_{i}$ of that neuron, namely:
\begin{spacing}{0.8}
\begin{center}
{\small{}
\begin{equation}
V_{i}\left(t+1\right)=\sum_{j=0}^{N-1}J_{i,j}A_{j}\left(t\right)+\mathfrak{I}_{i}+\sum_{j=0}^{N-1}\sigma_{i,j}^{\mathcal{N}}\mathcal{N}_{j}\left(t\right).\label{eq:change-of-variables}
\end{equation}
}
\par\end{center}{\small \par}
\end{spacing}

\noindent Since Eq.~(\ref{eq:activity-based-equations}) does not
depend on the variables $V_{i}\left(t\right)$, it can be solved without
knowing the behavior of the membrane potentials. For this reason,
the calculation of the probability distribution of $V_{i}\left(t\right)$
has always been neglected in the literature. Interestingly, we show
that it is possible to exactly derive the set of equations satisfied
by the membrane potentials, and that these equations provide a complementary
description of the network dynamics with respect to Eq.~(\ref{eq:activity-based-equations}).
Under the change of variables Eq.\textcolor{blue}{~}(\ref{eq:change-of-variables}),
we observe that Eq.\textcolor{blue}{~}(\ref{eq:activity-based-equations})
can be equivalently transformed into the following set of equations:
\begin{spacing}{0.8}
\begin{center}
{\small{}
\begin{equation}
V_{i}\left(t+1\right)=\sum_{j=0}^{N-1}J_{i,j}\mathscr{H}\left(V_{j}\left(t\right)-\theta_{j}\right)+\mathfrak{I}_{i}+\sum_{j=0}^{N-1}\sigma_{i,j}^{\mathcal{N}}\mathcal{N}_{j}\left(t\right),\quad i=0,...,N-1,\label{eq:voltage-based-equations}
\end{equation}
}
\par\end{center}{\small \par}
\end{spacing}

\noindent provided the matrix $J$ is invertible (note that this condition
can be eventually relaxed, see the supplementary information of \cite{Fasoli2018a}
for further details). Note also that Eqs.\textcolor{blue}{~}(\ref{eq:activity-based-equations})
and (\ref{eq:change-of-variables}) imply $A_{i}\left(t\right)=\mathscr{H}\left(V_{i}\left(t\right)-\theta_{i}\right)\in\left\{ 0,1\right\} $,
and that the binary output of a threshold unit can be loosely interpreted
as the firing rate of that neuron: $A_{i}\left(t\right)=0$ if the
$i$th neuron is not firing at time $t$, and $A_{i}\left(t\right)=1$
if it is firing at unit rate.

It is important to observe that the neural activities are discrete
random variables, therefore they are described by probability mass
functions (pmfs). We introduce the vector $\boldsymbol{A}=\left[\begin{array}[t]{ccc}
A_{0}, & \ldots, & A_{N-1}\end{array}\right]^{T}$ containing the activities of all $N$ neurons at time $t+1$, and
$\boldsymbol{A}'$ the vector of the activities of all neurons at
time $t$. We also define:
\begin{spacing}{0.8}
\begin{center}
{\footnotesize{}
\[
\Psi\overset{\mathrm{def}}{=}\left[\begin{array}{ccc}
\sigma_{0,0}^{\mathcal{N}} & \ldots & \sigma_{0,N-1}^{\mathcal{N}}\\
\vdots & \ddots & \vdots\\
\sigma_{N-1,0}^{\mathcal{N}} & \ldots & \sigma_{N-1,N-1}^{\mathcal{N}}
\end{array}\right],\quad\boldsymbol{\theta}\overset{\mathrm{def}}{=}\left[\begin{array}{c}
\theta_{0}\\
\vdots\\
\theta_{N-1}
\end{array}\right],\quad\boldsymbol{\mathfrak{I}}\overset{\mathrm{def}}{=}\left[\begin{array}{c}
\mathfrak{I}_{0}\\
\vdots\\
\mathfrak{I}_{N-1}
\end{array}\right],\quad\pmb{\mathscr{H}}\left(\boldsymbol{x}'-\boldsymbol{\theta}\right)\overset{\mathrm{def}}{=}\left[\begin{array}{c}
\mathscr{H}\left(x_{0}'-\theta_{0}\right)\\
\vdots\\
\mathscr{H}\left(x_{N-1}'-\theta_{N-1}\right)
\end{array}\right],
\]
}
\par\end{center}{\footnotesize \par}
\end{spacing}

\noindent where the matrix $\Psi$ is invertible by hypothesis. Moreover,
we introduce the matrix $\mathcal{C}=\left[\mathcal{C}_{i,j}\right]_{i,j=0,\cdots,2^{N}-2}$,
such that:
\begin{spacing}{0.8}
\begin{center}
{\small{}
\begin{equation}
\mathcal{C}_{i,j}=\delta_{i,j}+G_{i,2^{N}-1}-G_{i,j},\label{eq:matrix-C}
\end{equation}
}
\par\end{center}{\small \par}
\end{spacing}

\noindent where $\delta_{i,j}$ is the Kronecker delta, and:
\begin{spacing}{0.8}
\begin{center}
{\small{}
\begin{equation}
G_{i,j}\overset{\mathrm{def}}{=}\frac{1}{\left|\det\left(\Psi\right)\right|}\int_{\mathscr{V}_{i}^{\left(N\right)}}p_{\boldsymbol{\mathcal{N}}}\left(\Psi^{-1}\left[\boldsymbol{x}-J\pmb{\mathscr{B}}_{j}^{\left(N\right)}-\boldsymbol{\mathfrak{I}}\right]\right)d\boldsymbol{x},\quad i,j=0,\cdots,2^{N}-1.\label{eq:coefficients-G}
\end{equation}
}
\par\end{center}{\small \par}
\end{spacing}

\noindent In Eq.\textcolor{blue}{~}(\ref{eq:coefficients-G}), $\pmb{\mathscr{B}}_{i}^{\left(N\right)}$
is the $N\times1$ vector whose entries are the digits of the binary
representation of the index $i$ (e.g. $\pmb{\mathscr{B}}_{39}^{\left(6\right)}=\left[\begin{array}{cccccc}
1, & 0, & 0, & 1, & 1, & 1\end{array}\right]^{T}$). Moreover, the set $\mathscr{V}_{i}^{\left(N\right)}$ is defined
as follows:
\begin{spacing}{0.8}
\noindent \begin{center}
{\small{}
\[
\mathscr{V}_{i}^{\left(N\right)}\overset{\mathrm{def}}{=}\left\{ \boldsymbol{x}\in\mathbb{R}^{N}:\;i=\mathscr{D}\left(\pmb{\mathscr{H}}\left(\boldsymbol{x}-\boldsymbol{\theta}\right)\right)\right\} ,
\]
}
\par\end{center}{\small \par}
\end{spacing}

\noindent where $\mathscr{D}\left(\boldsymbol{\nu}\right)$ is the
decimal representation of the binary vector $\boldsymbol{\nu}$. For
example, for $N=2$, we get:{\small{}
\begin{align*}
\mathscr{V}_{0}^{\left(2\right)}= & \left\{ \left(x_{0},x_{1}\right)\in\mathbb{R}^{2}:x_{0}<\theta_{0},x_{1}<\theta_{1}\right\} \\
\\
\mathscr{V}_{1}^{\left(2\right)}= & \left\{ \left(x_{0},x_{1}\right)\in\mathbb{R}^{2}:x_{0}<\theta_{0},x_{1}\geq\theta_{1}\right\} \\
\\
\mathscr{V}_{2}^{\left(2\right)}= & \left\{ \left(x_{0},x_{1}\right)\in\mathbb{R}^{2}:x_{0}\geq\theta_{0},x_{1}<\theta_{1}\right\} \\
\\
\mathscr{V}_{3}^{\left(2\right)}= & \left\{ \left(x_{0},x_{1}\right)\in\mathbb{R}^{2}:x_{0}\geq\theta_{0},x_{1}\geq\theta_{1}\right\} .
\end{align*}
}{\small \par}

\noindent (note for example that, for any $\left(x_{0},x_{1}\right)\in\mathscr{V}_{3}^{\left(2\right)}$,
we get $\left[\begin{array}{c}
\mathscr{H}\left(x_{0}-\theta_{0}\right)\\
\mathscr{H}\left(x_{1}-\theta_{1}\right)
\end{array}\right]=\left[\begin{array}{c}
1\\
1
\end{array}\right]$, whose decimal representation is $3$). Finally, we define:
\begin{spacing}{0.8}
\begin{center}
{\small{}
\[
\widetilde{\boldsymbol{H}}=\left[\begin{array}{ccc}
H_{0}, & \ldots, & H_{2^{N}-2}\end{array}\right]^{T}\overset{\mathrm{def}}{=}\mathcal{C}^{-1}\widetilde{\boldsymbol{G}},\quad\widetilde{\boldsymbol{G}}\overset{\mathrm{def}}{=}\left[G_{i,2^{N}-1}\right]_{i=0,\ldots,2^{N}-2},\quad H_{2^{N}-1}\overset{\mathrm{def}}{=}1-\sum_{j=0}^{2^{N}-2}H_{j}.
\]
}
\par\end{center}{\small \par}
\end{spacing}

\noindent Then, in \cite{Fasoli2018a} we proved that the conditional
probability distribution of $\boldsymbol{A}$ given $\boldsymbol{A}'$,
and the stationary joint distribution of $\boldsymbol{A}$ in the
limit $t\rightarrow+\infty$, are:

\begin{spacing}{0.8}
\begin{center}
{\small{}
\begin{align}
P\left(\boldsymbol{A},t+1|\boldsymbol{A}',t\right)= & \frac{1}{\left|\det\left(\Psi\right)\right|}\int_{\mathscr{V}_{\mathscr{D}\left(\boldsymbol{A}\right)}^{\left(N\right)}}p_{\boldsymbol{\mathcal{N}}}\left(\Psi^{-1}\left[\boldsymbol{x}-J\boldsymbol{A}'-\boldsymbol{\mathfrak{I}}\right]\right)d\boldsymbol{x}\label{eq:neural-activity-conditional-probability-distribution}\\
\nonumber \\
\underset{t\rightarrow+\infty}{\lim}P\left(\boldsymbol{A},t\right)= & \frac{1}{\left|\det\left(\Psi\right)\right|}\sum_{j=0}^{2^{N}-1}H_{j}\int_{\mathscr{V}_{\mathscr{D}\left(\boldsymbol{A}\right)}^{\left(N\right)}}p_{\boldsymbol{\mathcal{N}}}\left(\Psi^{-1}\left[\boldsymbol{x}-J\pmb{\mathscr{B}}_{j}^{\left(N\right)}-\boldsymbol{\mathfrak{I}}\right]\right)d\boldsymbol{x}.\label{eq:neural-activity-joint-probability-distribution}
\end{align}
}
\par\end{center}{\small \par}
\end{spacing}

\noindent On the other hand, the membrane potentials $V_{i}\left(t\right)$
are continuous variables, therefore they are described by probability
density functions (pdfs). By introducing the vector $\boldsymbol{V}\overset{\mathrm{def}}{=}\left[\begin{array}[t]{ccc}
V_{0}, & \ldots, & V_{N-1}\end{array}\right]^{T}$, which contains the membrane potentials of all $N$ neurons at time
$t+1$, and the vector $\boldsymbol{V}'$ of the membrane potentials
of all neurons at time $t$, the conditional probability distribution
of $\boldsymbol{V}$ given $\boldsymbol{V}'$, and the stationary
joint distribution of $\boldsymbol{V}$ in the limit $t\rightarrow+\infty$,
are \cite{Fasoli2018a}:
\begin{spacing}{0.8}
\begin{center}
{\small{}
\begin{align}
p\left(\boldsymbol{V},t+1|\boldsymbol{V}',t\right)= & \frac{1}{\left|\det\left(\Psi\right)\right|}p_{\boldsymbol{\mathcal{N}}}\left(\Psi^{-1}\left[\boldsymbol{V}-J\pmb{\mathscr{H}}\left(\boldsymbol{V}'-\boldsymbol{\theta}\right)-\boldsymbol{\mathfrak{I}}\right]\right)\label{eq:membrane-potentials-conditional-probability-distribution}\\
\nonumber \\
\underset{t\rightarrow+\infty}{\lim}p\left(\boldsymbol{V},t\right)= & \frac{1}{\left|\det\left(\Psi\right)\right|}\sum_{j=0}^{2^{N}-1}H_{j}p_{\boldsymbol{\mathcal{N}}}\left(\Psi^{-1}\left[\boldsymbol{V}-J\pmb{\mathscr{B}}_{j}^{\left(N\right)}-\boldsymbol{\mathfrak{I}}\right]\right).\label{eq:membrane-potentials-joint-probability-distribution}
\end{align}
}
\par\end{center}{\small \par}
\end{spacing}

\noindent By comparing Eqs.\textcolor{blue}{~}(\ref{eq:neural-activity-conditional-probability-distribution})
and (\ref{eq:neural-activity-joint-probability-distribution}) with,
respectively, Eqs.\textcolor{blue}{~}(\ref{eq:membrane-potentials-conditional-probability-distribution})
and (\ref{eq:membrane-potentials-joint-probability-distribution}),
we observe that there exists a close relationship between the neural
activities $\boldsymbol{A}$ and the membrane potentials $\boldsymbol{V}$,
which we further investigate in SubSec.\textcolor{blue}{~}(\ref{subsec:Cross-neuron-correlations}).
It is also important to note that, generally, the integrals in Eqs.\textcolor{blue}{~}(\ref{eq:matrix-C})-(\ref{eq:neural-activity-joint-probability-distribution})
can be calculated only numerically or through analytical approximations.
However, in the specific case when the matrix $\Psi$ is diagonal,
and the stochastic variables $\mathcal{N}_{i}\left(t\right)$ are
independent, exact analytical solutions can be found in terms of the
cumulative distribution functions of the noise sources (see \cite{Fasoli2018a}).

\begin{figure}
\begin{centering}
\includegraphics[scale=0.22]{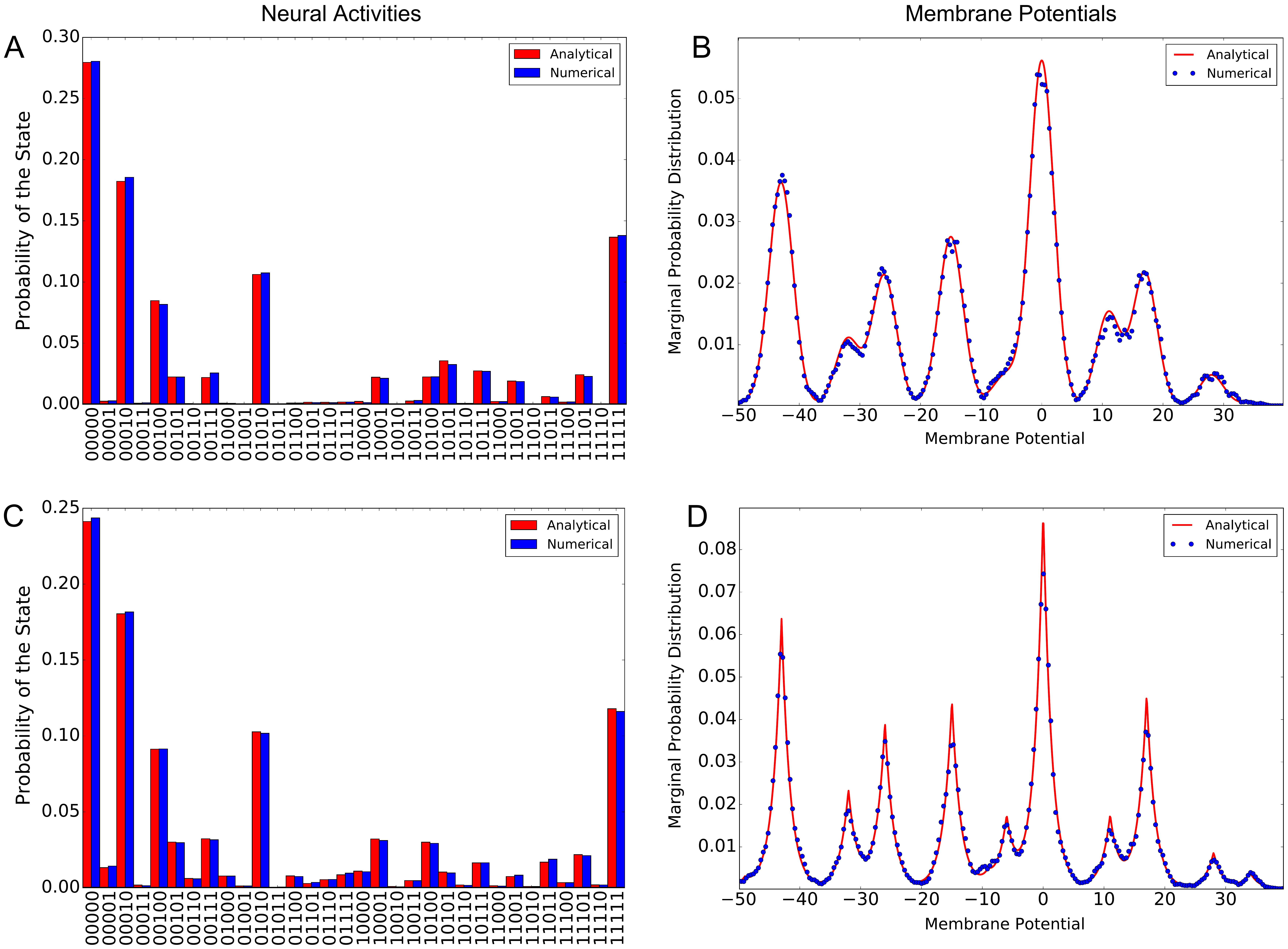}
\par\end{centering}
\caption{\label{fig:probability-distributions} \small\textbf{ Probability
distributions in stochastic networks.} This figure shows the probability
mass function $P\left(\boldsymbol{A},t\right)$ and the marginal probability
density function $p\left(V_{i},t\right)$ of the $0$th neuron, obtained
for the network parameters in Tab.\textcolor{blue}{~}(\ref{tab:Parameters-1}),
$\boldsymbol{\mathfrak{I}}=\left[\protect\begin{array}{ccccc}
0, & -3, & 2, & -4, & 0\protect\end{array}\right]^{T}$, $\Psi=\mathrm{diag}\left(\protect\begin{array}{ccccc}
2, & 3, & 2, & 3, & 3\protect\end{array}\right)$, and $t=100$. A) - B) Probability distributions obtained for correlated
normally-distributed noise sources, with uniform cross-correlation
(namely $\mathrm{Corr}\left(\mathcal{N}_{i}\left(t\right),\mathcal{N}_{j}\left(t\right)\right)=0.8$
and $\mathrm{Corr}\left(\mathcal{N}_{i}\left(t\right),\mathcal{N}_{j}\left(s\right)\right)=0$,
for $t\protect\neq s$ and $\forall i,\,j$). C) - D) Probability
distributions obtained for independent noise sources with Laplace
distributions, namely $p_{\boldsymbol{\mathcal{N}}}\left(\boldsymbol{x}\right)=\prod_{i=0}^{N-1}p_{\mathcal{N}_{i}}\left(x_{i}\right)$,
where $p_{\mathcal{N}_{i}}\left(x_{i}\right)=\frac{\sqrt{2}}{2}\exp\left(-\sqrt{2}\left|x_{i}\right|\right)$.
The red bars in panels A and C are calculated analytically from Eq.\textcolor{blue}{~}(\ref{eq:neural-activity-joint-probability-distribution}),
while the red curves in panels B and D are calculated according to
Eq.\textcolor{blue}{~}(\ref{eq:membrane-potentials-joint-probability-distribution}).
The blue bars in panels A, C, and the blue dots in panels B, D are
calculated numerically through a Monte Carlo method, namely by solving
Eq.\textcolor{blue}{~}(\ref{eq:activity-based-equations}) $10,000$
times, and then by calculating the probability distributions across
the repetitions.}
\end{figure}

In Fig.\textcolor{blue}{~}(\ref{fig:probability-distributions})
we show an example of the joint probability distributions $P\left(\boldsymbol{A},t\right)$
and $p\left(\boldsymbol{V},t\right)$ in the large-time limit, obtained
for the network parameters that we reported in Tab.\textcolor{blue}{~}(\ref{tab:Parameters-1}).
 In this figure we considered two distinct distributions of the noise
sources (namely correlated normally-distributed variables $\mathcal{N}_{i}\left(t\right)$,
as well as independent sources with Laplace distributions), and we
showed how they differently shape the probability distributions of
the neural activity and of the membrane potentials.

\subsection{Cross-neuron correlations \label{subsec:Cross-neuron-correlations}}

The joint probability distributions $P\left(\boldsymbol{A},t\right)$
and $p\left(\boldsymbol{V},t\right)$, that we reported in the previous
section, provide a complete probabilistic description of the network
in the limit $t\rightarrow+\infty$. In particular, these distributions
can be used for calculating cross-neuron correlations, which represents
a powerful tool for quantifying the exchange of information between
neurons. In \cite{Fasoli2018a}, we derived exact analytical expressions
of the Pearson correlation coefficient for $t\rightarrow+\infty$,
in the case when the matrix $\Psi$ is diagonal (i.e. $\Psi=\mathrm{diag}\left(\sigma_{0}^{\mathcal{N}},\ldots,\sigma_{N-1}^{\mathcal{N}}\right)$,
where $\sigma_{i}^{\mathcal{N}}\overset{\mathrm{def}}{=}\sigma_{i,i}^{\mathcal{N}}\;\forall i$),
while the noise sources are independent and normally distributed.
By applying Eq.\textcolor{blue}{~}(\ref{eq:neural-activity-joint-probability-distribution}),
we found that the pairwise correlation between the neural activities
of two neurons with indexes $i$ and $j$ is:
\begin{spacing}{0.80000000000000004}
\begin{center}
{\small{}
\begin{align}
\mathrm{Corr}\left(A_{i},A_{j}\right)= & \frac{\mathrm{Cov}\left(A_{i},A_{j}\right)}{\sqrt{\mathrm{Var}\left(A_{i}\right)\mathrm{Var}\left(A_{j}\right)}}\nonumber \\
\nonumber \\
\mathrm{Cov}\left(A_{i},A_{j}\right)= & \frac{1}{4}\sum_{n=0}^{2^{N}-1}H_{n}\left(1-2\overline{A}_{i}-E_{n,i}\right)\left(1-2\overline{A}_{j}-E_{n,j}\right)\nonumber \\
\nonumber \\
\mathrm{Var}\left(A_{i}\right)= & \overline{A}_{i}-\left(\overline{A}_{i}\right)^{2}\label{eq:correlation-between-the-neural-activities}\\
\nonumber \\
\overline{A}_{i}= & \frac{1}{2}\left(1-\sum_{n=0}^{2^{N}-1}H_{n}E_{n,i}\right)\nonumber \\
\nonumber \\
E_{n,i}= & \mathrm{erf}\left(\frac{\theta_{i}-\sum_{m=0}^{N-1}J_{i,m}\mathscr{B}_{n,m}^{\left(N\right)}-\mathfrak{I}_{i}}{\sqrt{2}\sigma_{i}^{\mathcal{N}}}\right).\nonumber 
\end{align}
}
\par\end{center}{\small \par}
\end{spacing}

\noindent In a similar way, from Eq.\textcolor{blue}{~}(\ref{eq:membrane-potentials-joint-probability-distribution})
we derived the following formula for the pairwise correlation between
the membrane potentials:
\begin{spacing}{0.80000000000000004}
\begin{center}
{\small{}
\begin{align}
\mathrm{Corr}\left(V_{i},V_{j}\right)= & \frac{\mathrm{Cov}\left(V_{i},V_{j}\right)}{\sqrt{\mathrm{Var}\left(V_{i}\right)\mathrm{Var}\left(V_{j}\right)}}\nonumber \\
\nonumber \\
\mathrm{Cov}\left(V_{i},V_{j}\right)= & \sum_{n=0}^{2^{N}-1}H_{n}\mathcal{R}_{n,i}^{\left(N\right)}\mathcal{R}_{n,j}^{\left(N\right)}\nonumber \\
\label{eq:correlation-between-the-membrane-potentials}\\
\mathrm{Var}\left(V_{i}\right)= & \left(\sigma_{i}^{\mathcal{N}}\right)^{2}+\sum_{n=0}^{2^{N}-1}H_{n}\left(\mathcal{R}_{n,i}^{\left(N\right)}\right)^{2}\nonumber \\
\nonumber \\
\mathcal{R}_{n,i}^{\left(N\right)}\overset{\mathrm{def}}{=} & \sum_{m=0}^{N-1}\left[\left(\mathscr{B}_{n,m}^{\left(N\right)}-\sum_{k=0}^{2^{N}-1}H_{k}\mathscr{B}_{k,m}^{\left(N\right)}\right)J_{i,m}\right].\nonumber 
\end{align}
}
\par\end{center}{\small \par}
\end{spacing}

\begin{figure}
\begin{centering}
\includegraphics[scale=0.22]{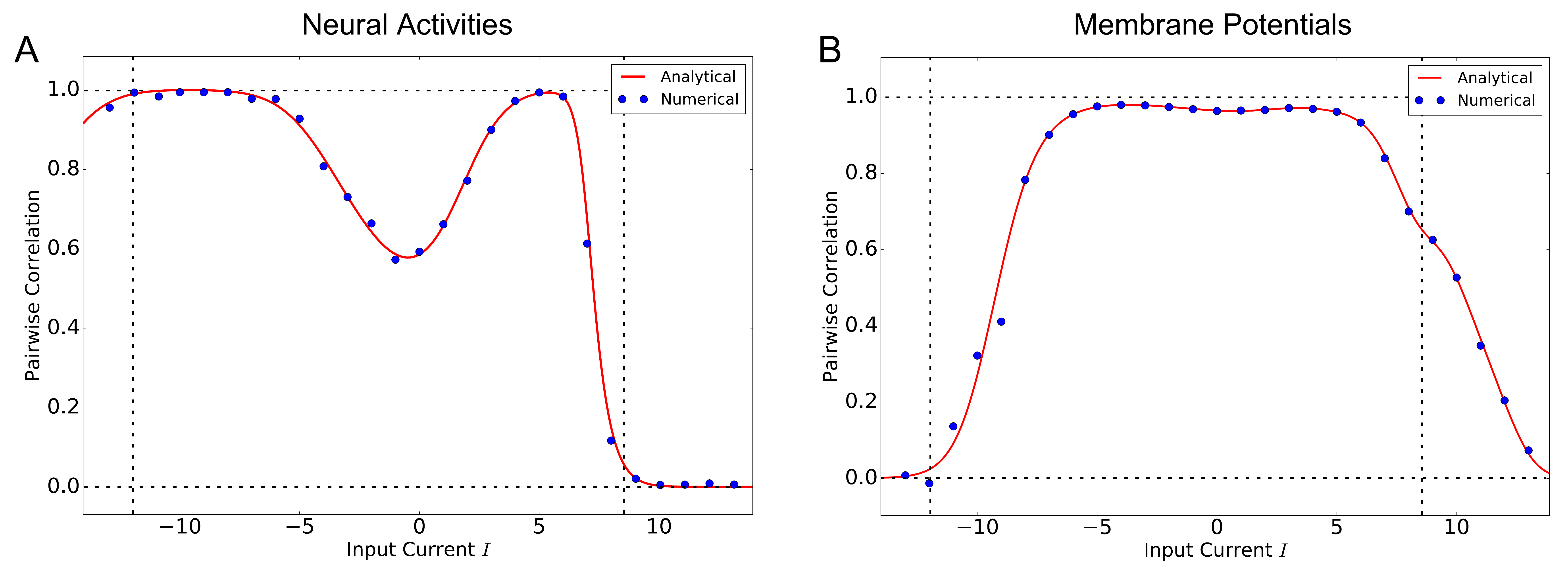}
\par\end{centering}
\caption{\label{fig:correlation} \small\textbf{ Cross-neuron correlations.}
This figure shows the dependence of the cross-neuron correlation of
the binary network on the external stimulus, in the specific case
of independent normally-distributed noise sources. Correlation is
calculated between the neurons with indexes $i=0$ and $i=4$, for
the values of the parameters reported in Tab.\textcolor{blue}{~}(\ref{tab:Parameters-1}),
$\boldsymbol{\mathfrak{I}}=I\left[\protect\begin{array}{ccccc}
1, & 1, & 1, & 1, & 1\protect\end{array}\right]^{T}$, $I\in\left[-14,14\right]$, $\Psi=\mathrm{diag}\left(\protect\begin{array}{ccccc}
2, & 3, & 2, & 3, & 3\protect\end{array}\right)$, and $t=100$. A) Correlation between neural activities. The red
curve is calculated analytically from Eq.\textcolor{blue}{~}(\ref{eq:correlation-between-the-neural-activities}).
B) Correlation between membrane potentials. The red curve is calculated
according to Eq.\textcolor{blue}{~}(\ref{eq:correlation-between-the-membrane-potentials}).
In both panels, the blue dots are calculated numerically through a
Monte Carlo method over $10,000$ repetitions. Moreover, the vertical
dashed lines correspond to the stimuli $I=-12$ and $I=8.5$. We chose
these values of the stimuli to show that low (respectively high) correlations
between neural activities do not necessarily correspond to low (respectively
high) correlations between the membrane potentials ($\mathrm{Corr}\left(A_{0},A_{4}\right)\approx0.99$
and $\mathrm{Corr}\left(V_{0},V_{4}\right)\approx0.02$ for $I=-12$,
while $\mathrm{Corr}\left(A_{0},A_{4}\right)\approx0.06$ and $\mathrm{Corr}\left(V_{0},V_{4}\right)\approx0.65$
for $I=8.5$).}
\end{figure}

\noindent In Fig.\textcolor{blue}{~}(\ref{fig:correlation}) we plotted
some examples of cross-correlations, obtained for the the network
parameters that we reported in Tab.\textcolor{blue}{~}(\ref{tab:Parameters-1}).
 This figure shows that variations of the external stimuli $\boldsymbol{\mathfrak{I}}$
switches the binary network between synchronous (i.e. highly correlated)
and asynchronous (i.e. uncorrelated) states. Moreover, we observe
that low (respectively high) correlations between neural activities
do not necessarily correspond to low (respectively high) correlations
between the membrane potentials. In other words, the linear relationship
between the neural activity and the membrane potentials, as given
by Eq.\textcolor{blue}{~}(\ref{eq:change-of-variables}), is not
reflected by the correlation structure of these variables. This result
proves that, despite the similarity of the corresponding equations
(which we already observed in SubSec.\textcolor{blue}{~}(\ref{subsec:Dualism-between-neural-activity-and-membrane-potentials}),
by comparing Eqs.\textcolor{blue}{~}(\ref{eq:neural-activity-conditional-probability-distribution})
and (\ref{eq:neural-activity-joint-probability-distribution}) with,
respectively, Eq.\textcolor{blue}{~}(\ref{eq:membrane-potentials-conditional-probability-distribution})
and (\ref{eq:membrane-potentials-joint-probability-distribution}))
and their linear relationship, neural activity and the membrane potentials
represent two opposed aspects of binary networks.

The interested reader is referred to \cite{Fasoli2018a} for a detailed
description of the conditions under which synchronous and asynchronous
states occur in the network, and for the extension of Eqs.\textcolor{blue}{~}(\ref{eq:correlation-between-the-neural-activities})
and (\ref{eq:correlation-between-the-membrane-potentials}) to encompass
higher-order (i.e. groupwise) correlations among an arbitrary number
of neurons.

\subsection{Pattern storage and retrieval in presence of noise \label{subsec:Pattern-storage-and-retrieval-in-presence-of-noise}}

In this section we consider the problem of storing $\mathcal{D}$
sequences of neural activity vectors $\boldsymbol{A}^{\left(i,0\right)}\rightarrow\boldsymbol{A}^{\left(i,1\right)}\rightarrow\cdots\rightarrow\boldsymbol{A}^{\left(i,\mathcal{T}_{i}\right)}$,
for $i=0,\ldots,\mathcal{D}-1$. In the context of content-addressable
memories, one aims to determine a synaptic connectivity matrix $J$
that stores these sequences in the binary network, so that each sequence
can be retrieved by initializing the network state to $\boldsymbol{A}\left(t=0\right)=\boldsymbol{A}^{\left(i,0\right)}$,
even in the presence of noise. Any method for calculating such a connectivity
matrix is typically called \textit{learning rule}.

Each transition $\boldsymbol{A}^{\left(i,n_{i}\right)}\rightarrow\boldsymbol{A}^{\left(i,n_{i}+1\right)}$
in the sequences is noise-resistant whenever \\ $P\left(\boldsymbol{A}^{\left(i,n_{i}+1\right)},t_{n_{i}}+1|\boldsymbol{A}^{\left(i,n_{i}\right)},t_{n_{i}}\right)\approx1$,
since under this condition the probability that the state $\boldsymbol{A}^{\left(i,n_{i}\right)}$
switches to a state other than $\boldsymbol{A}^{\left(i,n_{i}+1\right)}$
at the time instant $t_{n_{i}}+1$ is negligible. Therefore, according
to Eq.~(\ref{eq:neural-activity-conditional-probability-distribution}),
the sequences of neural activity can be stored in the network by solving
the following set of equations:
\begin{spacing}{0.8}
\begin{center}
{\footnotesize{}
\[
\int_{\mathscr{V}_{\mathscr{D}\left(\boldsymbol{A}\right)}^{\left(N\right)}}p_{\mathcal{N}}\left(\Psi^{-1}\left[\boldsymbol{x}-J\boldsymbol{A}'-\boldsymbol{\mathfrak{I}}\right]\right)d\boldsymbol{x}\approx\left|\det\left(\Psi\right)\right|,\quad\left(\boldsymbol{A}',\boldsymbol{A}\right)=\left(\boldsymbol{A}^{\left(i,n_{i}\right)},\boldsymbol{A}^{\left(i,n_{i}+1\right)}\right),\quad n_{i}=0,\ldots,\mathcal{T}_{i}-1,\quad i=0,\ldots,\mathcal{D}-1
\]
}
\par\end{center}{\footnotesize \par}
\end{spacing}

\noindent with respect to the connectivity matrix $J$. Generally,
these equations can be solved only numerically or through analytical
approximations. However, in the specific case when the matrix $\Psi$
is diagonal and the stochastic variables $\mathcal{N}_{i}\left(t\right)$
are independent, exact analytical solutions can be found.

In \cite{Fasoli2018a}, we considered the case of of independent normally-distributed
noise sources, and we found that, if the network is fully-connected
without self connections (so that $J_{i,i}=0$), the matrix $J$ that
stores the $\mathcal{D}$ neural sequences satisfies the following
sets of linear algebraic equations:
\begin{spacing}{0.8}
\begin{center}
{\small{}
\begin{equation}
\Omega^{\left(j\right)}\boldsymbol{J}^{\left(j\right)}=\boldsymbol{u}^{\left(j\right)},\quad j=0,\ldots,N-1.\label{eq:learning-rule}
\end{equation}
}
\par\end{center}{\small \par}
\end{spacing}

\noindent In Eq.~(\ref{eq:learning-rule}), $\boldsymbol{J}^{\left(j\right)}$
is the $\left(N-1\right)\times1$ vector with entries $J_{j,k}$ for
$k\neq j$. Moreover, if we define $\mathscr{T}\overset{\mathrm{def}}{=}\sum_{i=0}^{\mathcal{D}-1}\mathcal{T}_{i}$,
$\mathfrak{T}_{i}\overset{\mathrm{def}}{=}\sum_{k=1}^{i}\mathcal{T}_{k-1}$
(for $i>0$) and $\mathfrak{T}_{0}\overset{\mathrm{def}}{=}0$, then
$\boldsymbol{u}^{\left(j\right)}$ is a $\mathscr{T}\times1$ vector
with entries:
\begin{spacing}{0.8}
\begin{center}
{\small{}
\[
\left[\boldsymbol{u}^{\left(j\right)}\right]_{\mathfrak{T}_{i}+n_{i}}=\theta_{j}-\left(-1\right)^{A_{j}^{\left(i,n_{i}+1\right)}}K_{j}^{\left(i,n_{i}\right)}\sqrt{2}\sigma_{j}^{\mathcal{N}}-\mathfrak{I}_{j},\quad n_{i}=0,\ldots,\mathcal{T}_{i}-1,\quad i=0,\ldots,\mathcal{D}-1,
\]
}
\par\end{center}{\small \par}
\end{spacing}

\noindent where $K_{j}^{\left(i,n_{i}\right)}$ is any sufficiently
large and positive constant. Moreover, in Eq.~(\ref{eq:learning-rule}),
$\Omega^{\left(j\right)}$ is the $\mathscr{T}\times\left(N-1\right)$
matrix obtained by removing the $j$th column of the following matrix:
\begin{spacing}{0.80000000000000004}
\noindent \begin{center}
{\small{}
\[
\Omega=\left[\begin{array}{ccc}
A_{0}^{\left(0,0\right)} & \ldots & A_{N-1}^{\left(0,0\right)}\\
\vdots & \ddots & \vdots\\
A_{0}^{\left(0,\mathcal{T}_{0}-1\right)} & \ldots & A_{N-1}^{\left(0,\mathcal{T}_{0}-1\right)}\\
\vdots & \ddots & \vdots\\
A_{0}^{\left(\mathcal{D}-1,0\right)} & \ldots & A_{N-1}^{\left(\mathcal{D}-1,0\right)}\\
\vdots & \ddots & \vdots\\
A_{0}^{\left(\mathcal{D}-1,\mathcal{T}_{\mathcal{D}-1}-1\right)} & \ldots & A_{N-1}^{\left(\mathcal{D}-1,\mathcal{T}_{\mathcal{D}-1}-1\right)}
\end{array}\right].
\]
}
\par\end{center}{\small \par}
\end{spacing}

\noindent In particular, we observe that whenever $\boldsymbol{A}^{\left(i,0\right)}=\boldsymbol{A}^{\left(i,\mathcal{T}_{i}\right)}$,
the $i$th neural sequence is an oscillatory solution of Eq.~(\ref{eq:activity-based-equations})
with period $\mathcal{T}_{i}$, so that if the matrix $J$ is calculated
by solving Eq.~(\ref{eq:learning-rule}) and the network is initialized
to any state of the oscillation, the network will cycle repeatedly
through the same set of states. Moreover, in the special case $\mathcal{T}=1$,
the neural sequence represents a stationary solution of Eq.~(\ref{eq:activity-based-equations})
.

\begin{figure}
\begin{centering}
\includegraphics[scale=0.22]{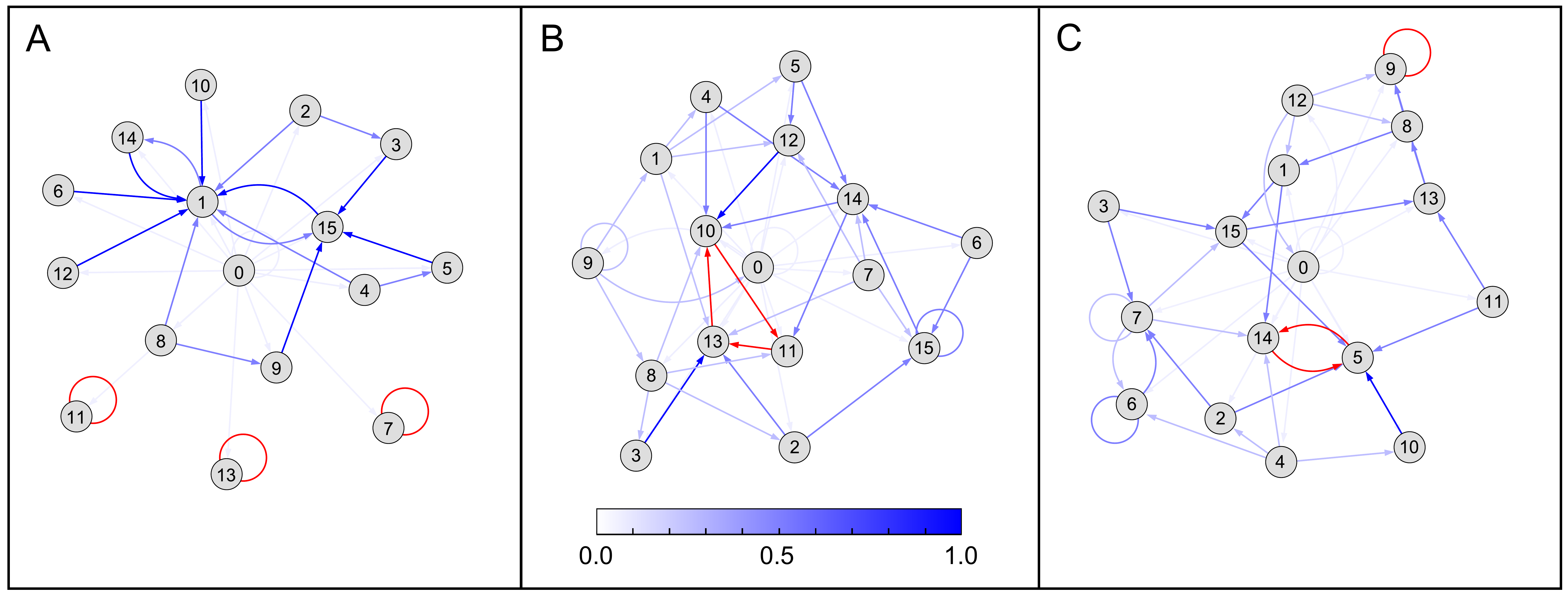}
\par\end{centering}
\caption{\label{fig:pattern-storage} \small\textbf{ Storage of activity patterns
and neural sequences.} This figure shows examples of activity patterns
and neural sequences (highlighted in red), stored in a stochastic
binary network, for $N=4$, $\boldsymbol{\mathfrak{I}}=\left[\protect\begin{array}{cccc}
0, & 0, & 0, & 0\protect\end{array}\right]^{T}$, and $\Psi=\mathrm{diag}\left(\protect\begin{array}{cccc}
50, & 50, & 50, & 50\protect\end{array}\right)$. The color gradation of the blue arrows is proportional to the magnitude
of $P\left(\boldsymbol{A}^{\left(i,n_{i}+1\right)},t_{n_{i}}+1|\boldsymbol{A}^{\left(i,n_{i}\right)},t_{n_{i}}\right)$
(see Eq.~(\ref{eq:neural-activity-conditional-probability-distribution})),
so that the arrows are white for every pair of states $\left(\boldsymbol{A}^{\left(i,n_{i}\right)},\boldsymbol{A}^{\left(i,n_{i}+1\right)}\right)$
such that $P\left(\boldsymbol{A}^{\left(i,n_{i}+1\right)},t_{n_{i}}+1|\boldsymbol{A}^{\left(i,n_{i}\right)},t_{n_{i}}\right)\approx0$,
while they are blue if the conditional probability is close to $1$.
The noise sources are supposed to be independent and normally distributed,
so that the synaptic connectivity matrix $J$ that stores the patterns
can be calculated according to Eq.~(\ref{eq:learning-rule}). In
these examples, we set $K_{j}^{\left(i,n_{i}\right)}=10\;\forall i,\:,j,\:,n_{i}$,
while the matrices $\Omega$ and the corresponding solutions $J$
are reported in Tab.~(\ref{tab:Parameters-2}). A) Storage of $3$
stationary states, highlighted in red. B) Storage of an oscillation
with period $\mathcal{T}=3$. C) Storage of a stationary state and
an oscillation with period $\mathcal{T}=2$. Note that the stationary
states and the oscillations are noise-resistant, despite the presence
of strong noise sources ($\sigma_{i}^{\mathcal{N}}=50\;\forall i$).}
\end{figure}

In Fig.~(\ref{fig:pattern-storage}), we show some examples of storage
of stationary patterns and oscillatory sequences with $\mathscr{T}=3$.
This figure is obtained for the the network parameters reported in
Tab.\textcolor{blue}{~}(\ref{tab:Parameters-2}), and proves that
the learning rule Eq.~(\ref{eq:learning-rule}) can be used to store
safely sequences of neural activity also in very noisy networks. 
\begin{table}
\begin{centering}
{\scriptsize{}}%
\begin{tabular}{|l|l|l|}
\hline 
\textbf{\small{}Panel A} & \textbf{\small{}Panel B} & \textbf{\small{}Panel C}\tabularnewline
\hline 
 &  & \tabularnewline
{\tiny{}$\Omega=\left[\begin{array}{cccc}
1 & 1 & 0 & 1\\
0 & 1 & 1 & 1\\
1 & 0 & 1 & 1
\end{array}\right]$} & {\tiny{}$\Omega=\left[\begin{array}{cccc}
1 & 0 & 1 & 1\\
1 & 1 & 0 & 1\\
1 & 0 & 1 & 0
\end{array}\right]$} & {\tiny{}$\Omega=\left[\begin{array}{cccc}
0 & 1 & 0 & 1\\
1 & 1 & 1 & 0\\
1 & 0 & 0 & 1
\end{array}\right]$}\tabularnewline
 &  & \tabularnewline
{\tiny{}$J\approx\left[\begin{array}{cccc}
0 & -1414 & -1414 & 2122\\
-1414 & 0 & -1414 & 2122\\
-1414 & -1414 & 0 & 2122\\
354 & 354 & 354 & 0
\end{array}\right]$} & {\tiny{}$J\approx\left[\begin{array}{cccc}
0 & 708 & 708 & 0\\
-2120 & 0 & 1414 & 1414\\
708 & 1414 & 0 & -1414\\
1 & -707 & 707 & 0
\end{array}\right]$} & {\tiny{}$J\approx\left[\begin{array}{cccc}
0 & 0 & -706 & 708\\
-1414 & 0 & 2122 & 708\\
-1060 & 354 & 0 & 354\\
708 & -706 & 706 & 0
\end{array}\right]$}\tabularnewline
 &  & \tabularnewline
\hline 
\end{tabular}
\par\end{centering}{\scriptsize \par}
\caption{\label{tab:Parameters-2} \textbf{Network parameters 2}. This table
reports the matrices $\Omega$ and the corresponding matrices $J$,
that we used for plotting Fig.~(\ref{fig:pattern-storage}).}
\end{table}

\section{Networks with quenched disorder \label{sec:Networks-with-quenched-disorder}}

The results reported in Secs.~ (\ref{sec:Analysis-of-bifurcations-in-deterministic-networks})
and (\ref{sec:Stochastic-networks}) are valid for binary neural networks
with arbitrary topology of the synaptic connections (which does not
evolve over time). For this reason, they can be applied to networks
with regular connectivity matrices $J$, as well as to random networks
with \textit{frozen} synaptic weights (see e.g. the network parameters
reported in Tab.~(\ref{tab:Parameters-1})). In other words, the
connectivity matrix of random networks can be interpreted as a single
realization of the synaptic wiring among neurons, generated according
to some known probability distribution $p_{J}$. These models are
said to present \textit{quenched disorder} \cite{Sherrington1976,Kirkpatrick1978,Hermann2012}.

Each realization of the connectivity matrix, generated according to
the distribution $p_{J}$, usually produces a distinct matrix $J$,
which in turn gives rise to distinct dynamical properties of the neural
activity. In particular, each realization typically produces distinct
bifurcation diagrams. For this reason, in order to obtain statistically
representative results, one needs to average the coordinates of the
bifurcation points over the variability of the matrix $J$. More generally,
one would be interested in determining the probability distribution
of the bifurcation points over the matrix $J$.

In \cite{Fasoli2018b} we derived semi-analytical expressions of these
probability distributions. For simplicity, we focused on the bifurcation
points of the stationary states in the zero-noise limit ($\sigma_{i,j}^{\mathcal{N}}\rightarrow0\;\forall i,j$
). We supposed that the entries $J_{i,j}$ of the connectivity matrix
can be decomposed as the product of a synaptic weight, $W_{i,j}$
(which represents the random strength of interaction of the $j$th
neuron on the $i$th neuron), with another random variable, $T_{i,j}$
(which represents either the presence, for $T_{i,j}=1$, or the absence,
for $T_{i,j}=0$, of a synaptic connections from the $j$th to the
$i$th neuron). The random variable $W_{i,j}$ is supposed to be continuous,
and distributed according to some distribution $p_{W_{i,j}}$. On
the other hand, the variable $T_{i,j}$ is discrete, and such that
$T_{i,j}=1$ with probability $\mathcal{P}_{i,j}\in\left[0,1\right]$,
while $T_{i,j}=0$ with probability $1-\mathcal{P}_{i,j}$. We also
supposed that the variables $\left\{ W_{i,j},T_{i,j}\right\} _{i,j=0,\cdots,N-1}$
are statistically independent. Because of these assumptions, the random
variable $\mathcal{I}_{i}$ (as given by Eq.~(\ref{eq:neural-sequence-stimulus-range});
note that the superscript $\left(j\right)$ can be omitted in the
case of stationary states studied in this section) is distributed
as follows:
\begin{spacing}{0.8}
\begin{center}
\textit{\small{}
\begin{align*}
p_{\mathcal{I}_{i}}\left(x\right)= & a_{i}\left(\theta_{i}-x\right)+b_{i}\delta\left(\theta_{i}-x\right)\\
\\
a_{i}\left(x\right)= & \sum_{\mathscr{S}\in\mathbb{P}\left(R\right)\backslash\emptyset}\left[\prod_{j\in\mathscr{S}}\mathcal{P}_{i,j}\right]\left[\prod_{j\in R\backslash\mathscr{S}}\left(1-\mathcal{P}_{i,j}\right)\right]\left[\left(\name_{j\in\mathscr{S}}p_{W_{i,j}}\right)\left(x\right)\right]\\
\\
b_{i}= & \prod_{j\in R}\left(1-\mathcal{P}_{i,j}\right),
\end{align*}
}
\par\end{center}{\small \par}
\end{spacing}

\noindent where $\mathbb{P}\left(R\right)$ represents the power set
of $R\overset{\mathrm{def}}{=}\left\{ i\in\left\{ 0,\cdots,N-1\right\} :\;A_{i}=1\right\} $.
Moreover, we call $F_{\mathcal{I}_{i}}$ the cumulative distribution
function of $\mathcal{I}_{i}$. Then, the coordinates of the bifurcation
points, $\Lambda_{\alpha}$ and $\Xi_{\alpha}$, are distributed as
follows:
\begin{spacing}{0.8}
\begin{center}
{\small{}
\begin{equation}
p_{X}\left(x\right)=p_{X^{c}}\left(x\right)+\sum_{q\in D}\left[F_{X}\left(x_{q}\right)-\underset{x\rightarrow x_{q}^{-}}{\lim}F_{X}\left(x\right)\right]\delta\left(x-x_{q}\right),\label{eq:probability-distribution-decomposition}
\end{equation}
}
\par\end{center}{\small \par}
\end{spacing}

\noindent for $X\in\left\{ \Lambda_{\alpha},\Xi_{\alpha}\right\} $.
In Eq.~(\ref{eq:probability-distribution-decomposition}), $\delta\left(\cdot\right)$
is the Dirac delta function, $p_{X^{c}}$ is the component of $p_{X}$
that describes the statistical behavior of the continuous values of
$X$, and $F_{X}$ is the cumulative distribution function of $X$.
Since, according to Eq.~(\ref{eq:neural-sequence-stimulus-range}),
$\Lambda_{\alpha}$ and $\Xi_{\alpha}$ are, respectively, the maximum
and minimum of the independent variables $\mathcal{I}_{i}$, they
must be distributed according to order statistics \cite{Vaughan1972,Bapat1989,Bapat1990,Hande1994}.
By calling $\mathrm{per}\left(\cdot\right)$ the matrix permanent,
in \cite{Fasoli2018b} we proved that $X^{c}$ is distributed as follows:
\begin{spacing}{0.8}
\begin{center}
{\small{}
\begin{align}
p_{\Lambda_{\alpha}^{c}}\left(x\right)= & \frac{1}{\left(\gamma_{\alpha,1}-1\right)!}\mathrm{per}\left(\left[\begin{array}{cc}
\boldsymbol{a}_{\alpha,1}\left(\boldsymbol{\theta}-\boldsymbol{x}\right), & \boldsymbol{F}_{\alpha,1}^{\left(\gamma_{\alpha,1}-1\right)}\left(x\right)\end{array}\right]\right)\nonumber \\
\label{eq:continuous-components}\\
p_{\Xi_{\alpha}^{c}}\left(x\right)= & \frac{1}{\left(\gamma_{\alpha,0}-1\right)!}\mathrm{per}\left(\left[\begin{array}{cc}
\boldsymbol{a}_{\alpha,0}\left(\boldsymbol{\theta}-\boldsymbol{x}\right), & \mathbb{I}_{\gamma_{\alpha,0},\gamma_{\alpha,0}-1}-\boldsymbol{F}_{\alpha,0}^{\left(\gamma_{\alpha,0}-1\right)}\left(x\right)\end{array}\right]\right),\nonumber 
\end{align}
}
\par\end{center}{\small \par}
\end{spacing}

\noindent where $\gamma_{\alpha,u}\overset{\mathrm{def}}{=}\left|\Gamma_{I_{\alpha},u}\right|$,
while $\left[\begin{array}{cc}
\boldsymbol{a}_{\alpha,1}\left(\boldsymbol{\theta}-\boldsymbol{x}\right), & \boldsymbol{F}_{\alpha,1}^{\left(\gamma_{\alpha,1}-1\right)}\left(x\right)\end{array}\right]$ and \\ $\left[\begin{array}{cc}
\boldsymbol{a}_{\alpha,0}\left(\boldsymbol{\theta}-\boldsymbol{x}\right), & \mathbb{I}_{\gamma_{\alpha,0},\gamma_{\alpha,0}-1}-\boldsymbol{F}_{\alpha,0}^{\left(\gamma_{\alpha,0}-1\right)}\left(x\right)\end{array}\right]$ are $\gamma_{\alpha,1}\times\gamma_{\alpha,1}$ and $\gamma_{\alpha,0}\times\gamma_{\alpha,0}$
matrices respectively, \\  $\boldsymbol{a}_{\alpha,u}\left(\boldsymbol{\theta}-\boldsymbol{x}\right)\overset{\mathrm{def}}{=}\left[a_{i}\left(\theta_{i}-x\right)\right]_{i\in\Gamma_{I_{\alpha},u}}$
and $\boldsymbol{F}_{\alpha,u}\left(x\right)\overset{\mathrm{def}}{=}\left[F_{\mathcal{I}_{i}}\left(x\right)\right]_{i\in\Gamma_{I_{\alpha},u}}$
are $\gamma_{\alpha,u}\times1$ column vectors, $\boldsymbol{F}_{\alpha,u}^{\left(v\right)}\left(x\right)\overset{\mathrm{def}}{=}\underset{v-\mathrm{times}}{\left[\underbrace{\begin{array}{ccc}
\boldsymbol{F}_{\alpha,u}\left(x\right), & \cdots, & \boldsymbol{F}_{\alpha,u}\left(x\right)\end{array}}\right]}$ is a $\gamma_{\alpha,u}\times v$ matrix, and $\mathbb{I}_{\gamma_{\alpha,0},\gamma_{\alpha,0}-1}$
is the $\gamma_{\alpha,0}\times\left(\gamma_{\alpha,0}-1\right)$
all-ones matrix. Moreover, in Eq.~(\ref{eq:probability-distribution-decomposition}),
$\left\{ x_{q}\right\} _{q\in D}$ represents the set of the discrete
values of $X$, at which the cumulative distribution function $F_{X}$,
namely:
\begin{spacing}{0.8}
\begin{center}
{\small{}
\begin{align}
F_{\Lambda_{\alpha}}\left(x\right)= & \frac{1}{\gamma_{\alpha,1}!}\mathrm{per}\left(\left[\boldsymbol{F}_{\alpha,1}^{\left(\gamma_{\alpha,1}\right)}\left(x\right)\right]\right)\nonumber \\
\label{eq:cumulative-distribution-functions}\\
F_{\Xi_{\alpha}}\left(x\right)= & \sum_{n=1}^{\gamma_{\alpha,0}}\frac{1}{n!\left(\gamma_{\alpha,0}-n\right)!}\mathrm{per}\left(\left[\begin{array}{cc}
\boldsymbol{F}_{\alpha,0}^{\left(n\right)}\left(x\right), & \mathbb{I}_{\gamma_{\alpha,0},\gamma_{\alpha,0}-n}-\boldsymbol{F}_{\alpha,0}^{\left(\gamma_{\alpha,0}-n\right)}\left(x\right)\end{array}\right]\right),\nonumber 
\end{align}
}
\par\end{center}{\small \par}
\end{spacing}

\noindent is (possibly) discontinuous. Note that $D=\Gamma_{I_{\alpha},1}$
and $D=\Gamma_{I_{\alpha},0}$, for $\Lambda_{\alpha}$ and $\Xi_{\alpha}$
respectively, while $x_{q}=\theta_{q}$.

The mean multistability diagram of the network is the plot of the
bifurcation points $\Lambda_{\alpha}$ and $\Xi_{\alpha}$, averaged
over the realizations of the synaptic connectivity matrix $J$. In
other words, the mean bifurcation points $\left\langle \Lambda_{\alpha}\right\rangle $
and $\left\langle \Xi_{\alpha}\right\rangle $ (where the brackets
$\left\langle \cdot\right\rangle $ represent the mean over the realizations)
correspond to the values of the stimulus $I_{\alpha}$ at which a
given neural activity state $\boldsymbol{A}$ loses its stability
on average, turning into another stationary state or an oscillation.
In \cite{Fasoli2018b} we proved that:
\begin{spacing}{0.8}
\begin{center}
{\small{}
\begin{equation}
\left\langle X\right\rangle =\int_{-\infty}^{+\infty}xp_{X^{c}}\left(x\right)dx+\sum_{q\in D}x_{q}\left[F_{X}\left(x_{q}\right)-\underset{x\rightarrow x_{q}^{-}}{\lim}F_{X}\left(x\right)\right]=\int_{-\infty}^{+\infty}\left[\mathscr{H}\left(x\right)-F_{X}\left(x\right)\right]dx,\label{eq:mean_bifurcation_points}
\end{equation}
}
\par\end{center}{\small \par}
\end{spacing}

\noindent where the functions $p_{X^{c}}\left(x\right)$ and $F_{X}\left(x\right)$
are given, respectively, by Eqs.~(\ref{eq:continuous-components})
and (\ref{eq:cumulative-distribution-functions}).

The probability that a given activity state $\boldsymbol{A}$ is stationary
for a fixed combination of stimuli $\widehat{\boldsymbol{I}}=\left[\begin{array}{ccc}
\widehat{I}_{0}, & \cdots, & \widehat{I}_{\mathfrak{P}-1}\end{array}\right]^{T}$, corresponds to the probability that $\widehat{\boldsymbol{I}}\in\mathfrak{V}$,
where the coordinates of the hyperrectangle $\mathfrak{V}$ are calculated
from Eq.~(\ref{eq:neural-sequence-stimulus-range}) for the given
state $\boldsymbol{A}$. In \cite{Fasoli2018b} we proved that this
probability can be calculated from the cumulative distribution of
the bifurcation points as follows:
\begin{spacing}{0.8}
\begin{center}
{\small{}
\begin{align}
 & P\left(\widehat{\boldsymbol{I}}\in\mathfrak{V}\right)=\prod_{\alpha=0}^{\mathfrak{P}-1}P\left(\widehat{I}_{\alpha}\in\mathcal{V}_{\alpha}\right)\label{eq:probability-for-fixed-stimuli}\\
\nonumber \\
 & P\left(\widehat{I}_{\alpha}\in\mathcal{V}_{\alpha}\right)=\begin{cases}
1-F_{\Xi_{\alpha}}\left(\widehat{I}_{\alpha}\right), & \mathrm{if}\;\;\Gamma_{I_{\alpha},1}=\emptyset\\
\\
F_{\Lambda_{\alpha}}\left(\widehat{I}_{\alpha}\right), & \mathrm{if}\;\;\Gamma_{I_{\alpha},0}=\emptyset\\
\\
F_{\Lambda_{\alpha}}\left(\widehat{I}_{\alpha}\right)\left[1-F_{\Xi_{\alpha}}\left(\widehat{I}_{\alpha}\right)\right], & \mathrm{otherwise}.
\end{cases}\nonumber 
\end{align}
}
\par\end{center}{\small \par}
\end{spacing}

\noindent Moreover, the probability to observe the state $\boldsymbol{A}$
in the whole multistability diagram of a single realization of the
matrix $J$ (i.e. the probability that $\boldsymbol{A}$ is stationary,
regardless of the specific combination of stimuli), corresponds to
the probability that the hyperrectangle $\mathfrak{V}$ has positive
hypervolume $\mathrm{vol}\left(\mathscr{V}\right)$. In \cite{Fasoli2018b}
we proved that this probability has the following expression:
\begin{spacing}{0.8}
\begin{center}
{\small{}
\begin{equation}
P\left(\mathrm{vol}\left(\mathfrak{V}\right)>0\right)=\prod_{\alpha=0}^{\mathfrak{P}-1}\left[\int_{-\infty}^{+\infty}p_{\Xi_{\alpha}^{c}}\left(x\right)F_{\Lambda_{\alpha}}\left(x\right)dx+\sum_{q\in\Gamma_{I_{\alpha},0}}\left[F_{\Xi_{\alpha}}\left(\theta_{q}\right)-\underset{x\rightarrow\theta_{q}^{-}}{\lim}F_{\Xi_{\alpha}}\left(x\right)\right]F_{\Lambda_{\alpha}}\left(\theta_{q}\right)\right].\label{eq:probability-regardless-of-the-stimuli}
\end{equation}
}
\par\end{center}{\small \par}
\end{spacing}

\noindent Eqs.~(\ref{eq:probability-distribution-decomposition})-(\ref{eq:probability-regardless-of-the-stimuli})
provide a complete description of the statistical properties of the
stationary states in networks with quenched disorder. It is also important
to note that these equations are semi-analytical, since they are expressed
in terms of 1D integrals containing the distribution $p_{W_{i,j}}$.
These integrals may be calculated exactly for some $p_{W_{i,j}}$,
for example in the case of normally-distributed weights. However,
for simplicity, in this review and in \cite{Fasoli2018b}, they are
calculated through numerical integration schemes, because fully-analytical
expressions may be very cumbersome.

\begin{figure}
\begin{centering}
\includegraphics[scale=0.25]{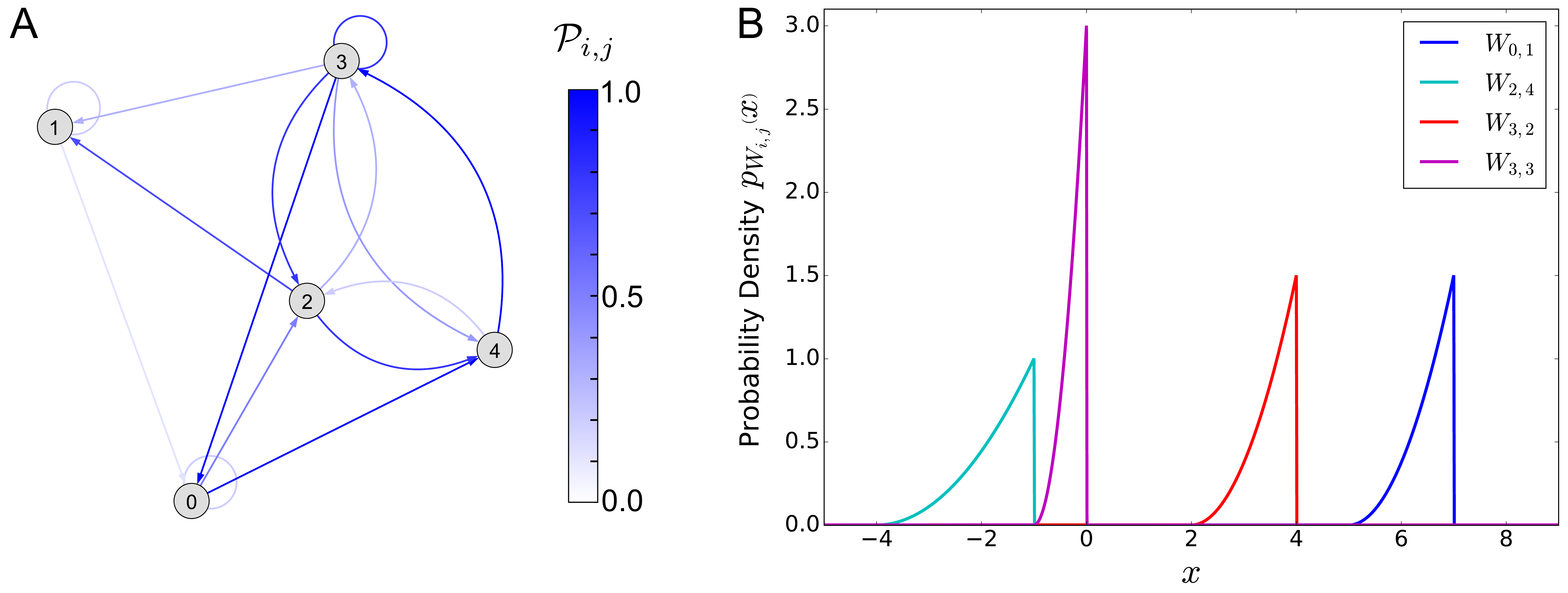}
\par\end{centering}
\caption{\label{fig:quenched-disorder-synaptic-probability-distribution} \small\textbf{
Probability distribution of the synaptic connections.} This figure
reports the probability distribution of the network topology and of
the synaptic weights, given the values of the network parameters reported
in Tab.~(\ref{tab:Parameters-3}). A) Probability distribution of
the network topology, namely the graph of the matrix $\left[\mathcal{P}_{i,j}\right]_{i,j=0,\cdots,N-1}$.
The $N=5$ nodes in the graph represent the neurons in the network,
while the arrow from the $j$th to the $i$th neuron represents the
probability to observe the $i\leftarrow j$ synaptic connection in
a single realization of the matrix $J$ (note that the color gradation
is proportional to $\mathcal{P}_{i,j}$). B) Examples of the powerlaw
probability distributions of the synaptic weights $W_{i,j}$, see
Eq.~(\ref{eq:powerlaw-distribution}).}
\end{figure}

\begin{figure}
\begin{centering}
\includegraphics[scale=0.22]{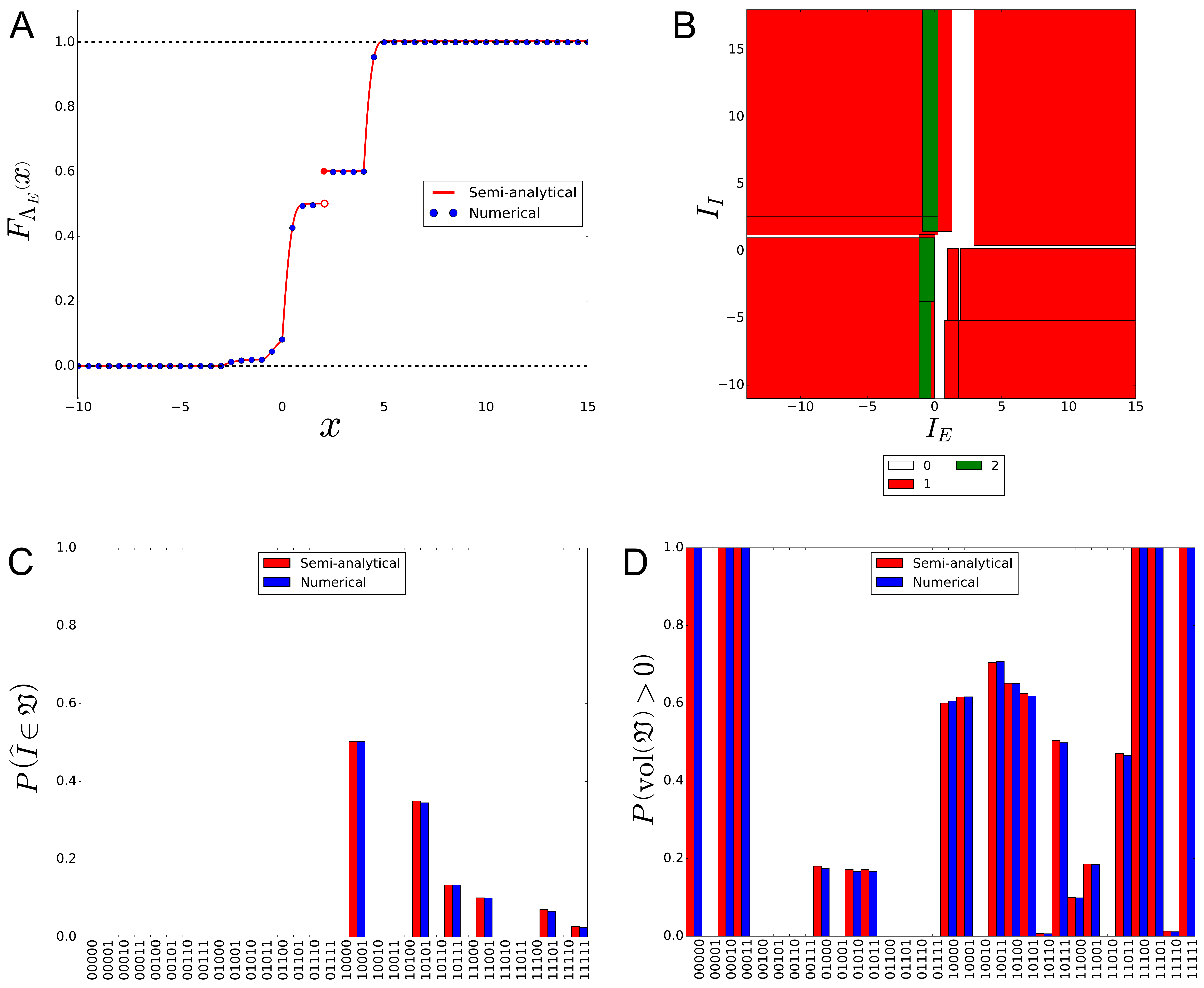}
\par\end{centering}
\caption{\label{fig:quenched-disorder-results} \small\textbf{ Stationary
behavior of a binary network with quenched disorder.} This figure
reports the probabilistic properties of the stationary states of a
binary network with quenched disorder. The matrix $J$ is generated
randomly from the powerlaw distribution Eq.~(\ref{eq:powerlaw-distribution}),
and for the network parameters in Tab.~(\ref{tab:Parameters-3})
(see also Fig.~(\ref{fig:quenched-disorder-synaptic-probability-distribution})).
A) Cumulative distribution function $F_{\Lambda_{E}}\left(x\right)$
of the activity state $\boldsymbol{A}=\left[\protect\begin{array}{ccccc}
1, & 0, & 1, & 1, & 0\protect\end{array}\right]^{T}$. The red curve is derived semi-analytically from Eq.~(\ref{eq:cumulative-distribution-functions}),
while the blue dots are calculated numerically through a Monte Carlo
method over $10,000$ repetitions of the synaptic connectivity matrix.
B) Mean multistability diagram of the network, obtained semi-analytically
from Eq.~(\ref{eq:mean_bifurcation_points}). C) -D) Occurrence probability
of the stationary states, obtained for the fixed stimuli $\left[\protect\begin{array}{c}
I_{E}\protect\\
I_{I}
\protect\end{array}\right]=\widehat{\boldsymbol{I}}=\left[\protect\begin{array}{c}
1\protect\\
0
\protect\end{array}\right]$ (panel C), and regardless of the stimuli (panel D). The red bars
in panel C (respectively, panel D) are derived semi-analytically from
Eq.~(\ref{eq:probability-for-fixed-stimuli}) (respectively, Eq.~(\ref{eq:probability-regardless-of-the-stimuli})),
while the blue bars are calculated numerically through a Monte Carlo
method (see \cite{Fasoli2018b} for more details).}
\end{figure}

In Figs.~(\ref{fig:quenched-disorder-synaptic-probability-distribution})
and (\ref{fig:quenched-disorder-results}) we show an example of these
results for a specific distribution of the connectivity matrix. 
 We consider a network composed of $3$ excitatory neurons (with indexes
$i=0,1,2$), and $2$ inhibitory neurons ($i=3,4$). The excitatory
and inhibitory neurons receive, respectively, external stimuli $I_{E}$
and $I_{I}$. We also assume that the synaptic weights $W_{i,j}$
are distributed according to the following powerlaw distribution:
\begin{spacing}{0.8}
\begin{center}
{\small{}
\begin{equation}
p_{W_{i,j}}\left(x\right)=\begin{cases}
\frac{3}{\mathfrak{W}}\left(\frac{x-\mathfrak{S}}{\mathfrak{W}}\right)^{2}, & \mathrm{if}\;\;\mathfrak{S}\leq x\leq\mathfrak{S}+\mathfrak{W}\\
\\
0, & \mathrm{otherwise},
\end{cases}\label{eq:powerlaw-distribution}
\end{equation}
}
\par\end{center}{\small \par}
\end{spacing}

\noindent where $\mathfrak{S}$ and $\mathfrak{W}$ represent, respectively,
the horizontal shift and the width of the support of the distribution.
To conclude, in Tab.~(\ref{tab:Parameters-3}) we reported the values
of the parameters $\mathcal{P}$, $\boldsymbol{\theta}$, $\mathfrak{S}$
and $\mathfrak{W}$ that we chose for this network. 
\begin{table}
\begin{centering}
\textbf{\small{}}%
\begin{tabular}{|lllll|}
\hline 
 &  &  &  & \tabularnewline
 & {\small{}$\mathcal{P}=\left[\begin{array}{ccccc}
0.2 & 0.1 & 0 & 1 & 0\\
0 & 0.2 & 0.7 & 0.3 & 0\\
0.5 & 0 & 0 & 0.8 & 0.2\\
0 & 0 & 0.3 & 0.8 & 1\\
1 & 0 & 0.8 & 0.4 & 0
\end{array}\right],$} &  & {\small{}$\boldsymbol{\theta}=\left[\begin{array}{c}
0\\
3\\
2\\
1\\
2
\end{array}\right]$} & \tabularnewline
 &  &  &  & \tabularnewline
 & {\small{}$\mathfrak{S}=\left[\begin{array}{ccccc}
5 & 5 & \times & -1 & \times\\
\times & 3 & 1 & -2 & \times\\
3 & \times & \times & -3 & -4\\
\times & \times & 2 & -1 & -1\\
5 & \times & 1 & -3 & \times
\end{array}\right],$} &  & {\small{}$\mathfrak{W}=\left[\begin{array}{ccccc}
1 & 2 & \times & 1 & \times\\
\times & 1 & 1 & 2 & \times\\
2 & \times & \times & 1 & 3\\
\times & \times & 2 & 1 & 1\\
1 & \times & 1 & 2 & \times
\end{array}\right]$} & \tabularnewline
 &  &  &  & \tabularnewline
\hline 
\end{tabular}
\par\end{centering}{\small \par}
\caption{\label{tab:Parameters-3} \textbf{Network parameters 3}. This table
contains the values of the parameters that we used for plotting Figs.~(\ref{fig:quenched-disorder-synaptic-probability-distribution})
and (\ref{fig:quenched-disorder-results}) The symbol $\times$ in
the matrices $\mathfrak{S}$ and $\mathfrak{W}$ means that the probability
distributions of the stationary states and of the bifurcation points
are not affected by those parameters, since the corresponding synaptic
connections are absent ($\mathcal{P}_{i,j}=0$).}
\end{table}
 Fig.~(\ref{fig:quenched-disorder-synaptic-probability-distribution})
reports the graph of the matrix $\left[\mathcal{P}_{i,j}\right]_{i,j=0,\cdots,N-1}$
and some examples of the powerlaw distribution of the sinaptic weights
$W_{i,j}$ (see Eq.~(\ref{eq:powerlaw-distribution})). Moreover,
in Fig.~(\ref{fig:quenched-disorder-results}) we show the the mean
multistability diagram of the network, as well as the occurrence probability
of the activity states for fixed stimuli (i.e. $I_{E}=1$ and $I_{I}=0$)
and regardless of the stimuli (see, respectively, Eqs.~(\ref{eq:mean_bifurcation_points}),
(\ref{eq:probability-for-fixed-stimuli}) and (\ref{eq:probability-regardless-of-the-stimuli})).

\section{Discussion \label{sec:Discussion}}

New mathematical techniques for analytically investigating finite-size
and small-size neural network models are invaluable theoretical tools
for studying the brain at its multiple scales of spatial organization,
that complement the already existing mean-field approaches. Studying
how the complexity and dynamics of neuronal network activity change
with the network size is of fundamental importance to understand why
networks in the brain appear organized at multiple spatial scales.
In this article, we reviewed the effort we made in this direction,
trying to fill the gap in the current neuro-mathematical literature.
In the following, we discuss strengths and weaknesses of the approach
we developed so far, the implications of our work for specific issues
related to bifurcation dynamics and learning, and for future progress
in the understanding of the function of networks in the brain.

\subsection{Advantages and weakness of our approach \label{subsec:Advantages-and-weakness-of-our-approach}}

\subsubsection{Bifurcation analysis of binary networks \label{subsec:Bifurcation-analysis-of-binary-networks}}

An effective tool for studying spin networks in physics is represented
by their energy (Hamiltonian) function. In order to study the low-temperature
physical properties of the network at the thermodynamic equilibrium,
one is often interested in finding out the global (and possibly degenerate)
minimum energy state of the network. This is known as the \textit{ground
state} of the system, and it can be calculated by minimizing the energy
function over the space of all possible spin configurations at absolute
temperature. This is an optimization problem, which, for networks
on non-planar or three- or higher-dimensional lattices, has been proven
to be NP-hard \cite{Papadimitriou1982}. 

In the study of discrete-time binary neural networks, energy (or,
more generally, Lyapunov) functions typically are known only for asynchronously
updated neurons with symmetric synaptic connections \cite{Hopfield1982}.
For neural networks with asymmetric connectivity matrices and/or synchronous
update, like the one we considered in this review, the search for
the ground state(s) turns into the more general problem of determining
the long-time (non-equilibrium) states in the zero-noise limit. It
is important to observe that this problem is even more formidable
than the search for the ground states, due to the intractable number
of oscillatory sequences that can eventually be observed during the
network dynamics. We are not aware of any algorithm that performs
efficiently this highly demanding combinatorial analysis in synchronously-updated
networks with asymmetric connections. The algorithm that we introduced
in \cite{Fasoli2019} represents an attempt to tackle this problem,
in the specific case of networks with sparse synaptic connections.

Once the set all the possible stationary and oscillatory solutions
has been evaluated, Eq.~(\ref{eq:neural-sequence-stimulus-range})
provides a fast, analytical way to calculate the bifurcation diagram
of the network. On the other hand, the bifurcation analysis of networks
composed of neurons with graded output typically requires numerical
continuation techniques \cite{Kuznetsov1998}, which do not provide
any analytical intuition of the mechanisms underlying the changes
of dynamics.

\subsubsection{Probability distributions and cross-neuron correlations \label{subsec:Probability-distributions-and-cross-neuron-correlations}}

The analytical results reported in SubSecs.~(\ref{subsec:Dualism-between-neural-activity-and-membrane-potentials})
and (\ref{subsec:Cross-neuron-correlations}) provide a complete description
of the probabilistic behavior of the neural activity and of the membrane
potentials in the long-time regime of single network realizations.
These results provide new \textit{qualitative} insights into the mechanisms
underlying stochastic neuronal dynamics, which hold for any network
size, and therefore are not restricted to small networks only.

However, the main drawback of Eqs.~(\ref{eq:neural-activity-joint-probability-distribution})
and (\ref{eq:membrane-potentials-joint-probability-distribution})
(and, as a consequence, also of Eqs.~(\ref{eq:correlation-between-the-neural-activities})
and (\ref{eq:correlation-between-the-membrane-potentials})), is represented
by the \textit{quantitative} evaluation of the probability distributions
and of the cross correlations. This requires the calculation of a
number of coefficients $H_{i}$ that increases exponentially with
the network size, and therefore proves intractable already for networks
composed of a few tens of neurons. A more efficient quantitative estimation
of these quantities for large networks can be performed numerically
through Monte Carlo methods.

In order to provide a complete description of the stationary behavior
of networks with quenched disorder, the calculation of the probability
distributions of the bifurcation points across network realizations,
that we reported in Sec.~(\ref{sec:Networks-with-quenched-disorder})
(see Eqs.~(\ref{eq:probability-distribution-decomposition})-(\ref{eq:mean_bifurcation_points})),
as well as the calculation of the occurrence probability of the stationary
states (Eqs.~(\ref{eq:probability-for-fixed-stimuli}) and (\ref{eq:probability-regardless-of-the-stimuli})),
must be performed for every stationary state of the model. Unfortunately,
these states are not known a priory, therefore the calculation of
the probability distributions must be repeated for all the $2^{N}$
combinations of the neural activity states. A possible solution to
this problem, in the specific case of sparse networks, is represented
by the sparse-efficient algorithm that we introduced in \cite{Fasoli2019}.
This algorithm allows a fast evaluation of the stationary solutions
of the network, so that the calculation of the probability distributions
can be performed only on the actual stationary states detected by
the sparse-efficient algorithm.

To conclude, another disadvantage of Eqs.~(\ref{eq:probability-distribution-decomposition})-(\ref{eq:probability-regardless-of-the-stimuli})
is represented by the numerical calculation of the matrix permanent,
which is computationally demanding. The fastest known technique for
calculating the permanent of arbitrary matrices is the \textit{Balasubramanian-Bax-Franklin-Glynn
(BBFG) formula} \cite{Balasubramanian1980,Bax1996,Bax1998,Glynn2010},
which has complexity $O\left(2^{N-1}N^{2}\right)$. In order to alleviate
this computational bottleneck, in \cite{Fasoli2018b} we derived a
closed-form analytical expression of the permanent of uniform block
matrices, which proved much faster than the BBFG formula. This solution
allowed us to speed up considerably the calculation of the bifurcation
structure of statistically-homogeneous multi-population networks with
quenched disorder, which are often considered to be a good approximation
of biologically realistic circuits, see e.g. \cite{Faugeras2009,Hermann2012,Cabana2013}.

\subsubsection{Learning rule \label{subsec:Learning-rule}}

The algorithm we introduced in SubSec.~(\ref{subsec:Pattern-storage-and-retrieval-in-presence-of-noise})
for storing some desired sequences of neural activity, was obtained
in \cite{Fasoli2018a} by manipulating the conditional probability
distribution of the neural activity (see Eq.~(\ref{eq:neural-activity-conditional-probability-distribution})).
This distribution does not depend on the coefficients $H_{i}$, therefore
the learning rule has polynomial complexity. For this reason, its
applicability is not restricted to small networks only.

Another interesting property of this learning rule is represented
by the possibility to store sequences of neural activity also in noisy
networks. Typically, noise can break a neural sequence if the stochastic
fluctuations are sufficiently strong. However, our algorithm is designed
for being noise-resistant, namely the probability of breaking the
sequence under the influence of noise can be made arbitrarily small.

In the literature, several learning rules have been proposed for networks
of binary neurons. A mechanism for storing and retrieving static patterns
of neural activity in networks with symmetric connectivity was proposed
by Hopfield \cite{Hopfield1982}. The storage of sequences of temporally
evolving patterns was investigated by Sompolinsky, Kanter and Kleinfeld
\cite{Sompolinsky1986,Kleinfeld1986}, for networks with non-instantaneous
synaptic transmission between neurons. In \cite{Dehaene1987}, Dehaene
et al introduced an alternative approach based on temporally evolving
synapses. Then, Buhmann and Schulten \cite{Buhmann1987} showed that
asymmetric connectivity and noise are sufficient conditions for storing
and retrieving temporal sequences, without further assumptions on
the biophysical properties of the synaptic connections.

It is important to observe that the learning rule introduced in SubSec.~(\ref{subsec:Pattern-storage-and-retrieval-in-presence-of-noise})
does not require transmission delays, time-dependent synaptic strengths,
or the presence of noise. Only asymmetric synaptic connections are
required. This is compatible with experimental observation, in that
the vast majority of synapses in real biological networks are asymmetric
\cite{DeFelipe1999}. Our result can be considered as an extension
to stochastic networks of the associating learning rule for deterministic
models, introduced by Personnaz et al in \cite{Personnaz1986}.

\subsection{Open problems and future directions for mathematical developments
\label{subsec:Open-problems-and-future-directions-for-mathematical-developments}}

While in \cite{Fasoli2019} we proposed an efficient solution for
performing the bifurcation analysis of binary networks with sparse
connectivity, fast algorithms for dense networks proved more difficult
to develop. In \cite{Fasoli2019} we showed that, in the specific
case of homogeneous networks with regular topologies, it is possible
to take advantage of the symmetries of the network equations to speed
up the calculation of the bifurcation diagram. This observation allowed
us to introduce an algorithm which runs in linear time with respect
to the network size. However, the development of efficient algorithms
for dense networks with arbitrary topology of the synaptic connections
represents a much more difficult challenge, and it still remains an
open problem to be addressed in future work.

Another open problem is represented by the bifurcation analysis of
large and medium-size networks. Since computer's processing power
increases over time, the size of the networks that can be studied
through combinatorial approaches, such as the one we introduced in
\cite{Fasoli2019}, is expected to increase accordingly. For this
reason, we believe that these methods are key to the future development
of computational neuroscience and of the physics of complex systems.
Beyond brute-force processing power, other techniques can be developed
for accelerating combinatorial algorithms; in particular, the algorithm
that we developed in \cite{Fasoli2019} lends itself to be parallelized
over several processors. It also is important to note that our algorithm
calculates exact bifurcation diagrams, while the calculation of approximate
bifurcation diagrams, through a heuristic search of the oscillatory
and stationary solutions of the network equations, would prove much
faster.

To conclude, an important open problem in the study of networks with
quenched disorder, is represented by models with correlated synaptic
weights. In Sec.~(\ref{sec:Networks-with-quenched-disorder}), we
calculated the probability distribution of the bifurcation points,
through the results derived in \cite{Vaughan1972,Bapat1989,Bapat1990,Hande1994}
for the order statistics of a set of independent random variables.
On the other hand, it is known that synaptic weights in real cortical
circuits are correlated, as a consequence, for example, of synaptic
plasticity mechanisms. However, a generalization of order statistics
to sets of arbitrarily correlated random variables is still out of
reach. More generally, the analytical investigation of networks with
correlated synaptic weights has proven a formidable problem in mathematical
neuroscience, that has challenged also the mean-field theories of
large-size systems. Because of its biological relevance, the presence
of synaptic correlations represents an important ingredient in the
study of neural network dynamics, which needs to be addressed in future
research for increasing the biological plausibility of the models. 

\subsection{Possible implications of this mathematical progress for neuroscience
\label{subsec:Possible-implications-of-this-mathematical-progress-for-neuroscience} }

Being able to understand analytically the dynamics of finite-size
networks from the circuit's equations is potentially important for
improving our understanding of how neural circuits work and of how
and why their function is impaired in certain neural disorders. The
pattern and power of the external inputs, the pattern of anatomical
connectivity, the synaptic strength and the relative firing rate of
excitatory and inhibitory neurons, are all key elements in determining
the functional organization and output of a circuit. Yet, it is not
known how exactly these factor combine to produce brain functions
and to cause dysfunctions. Mathematical work to understanding neural
networks of arbitrary size could be useful to address two questions
relevant to these issues. 

The first question regards the relationship between anatomy, functional
coupling and population coding in local neural circuits. Our own work
(e.g. \cite{Panzeri1999,Panzeri2015,Zuo2015,Runyan2017}), and that
of many others \cite{Zohary1994,Singer1999,Ecker2010,Cohen2011,Brette2012,Moreno-Bote2014},
has shown that both the dynamics of individual neurons and of populations,
and the functional coupling between cells, is crucial for shaping
information in population codes and for behavior. Functional coupling
between cells may arise both because of anatomical connectivity between
neurons, but also because of other factors such as common inputs.
Thus, the relationship between functional coupling, circuit's anatomy,
and population-level information coding has remained largely unaddressed.
However, recent advances in experimental techniques allow the simultaneous
functional imaging of activity of several neurons in mice during sensory
or cognitive tasks, as well as the post-mortem measure by Electron-Microscopy
of the anatomical connectivity of the same set of neurons that were
functionally imaged in vivo \cite{Lee2016}. The mathematical work
reviewed here develops a set of tools that could be used to complement
the measure of anatomy and physiology from the same circuits, and
help bridging the gap between these two measures. Our tools, when
coupled with modern experimental techniques such as those described
above, could be used, in particular, to understand what is the consequence
of specific patterns of recurrent anatomical connectivity on population
coding and circuit dynamics, and then to test these theoretical relationships
on real data. 

A second possible direction of relevance for neuroscience of our work
is to use these tools to understand the neural origin of certain brain
disorders. For example, it is thought that Autism Spectrum Disorders
(ASD) result, at least in part, from abnormal changes in the functional
organization and dynamics of neural circuits \cite{Rubenstein2003,Zoghbi2003,Sahin2015,Ip2018}.
However, although many changes in parameters such as the strength
of synaptic connections and/or the change in firing properties of
certain classes of neurons have been observed in ASD, it is still
unclear how different elementary changes in neural parameters combine
to change the circuit's function. Being able to test directly, and
understand mathematically at a deep level, how the changes of such
basic neural properties affect the circuit's dynamics at different
spatial scales, including scales that involve finite-size neural networks,
could be useful to understand the origin and consequences of aberrant
circuit function in ASD conditions, as well as in other neural disorders. 

\section*{Acknowledgments}

This research was in part supported by the Simons Foundation (SFARI,
Explorer Grant No. 602849) and by the BRAIN Initiative (Grant No.
R01 NS108410). The funders had no role in study design, data collection
and analysis, decision to publish, interpretation of results, or preparation
of the manuscript.

\bibliographystyle{plain}
\bibliography{Bibliography}

\end{document}